\documentclass[prb,twocolumn,superscriptaddress,showpacs]{revtex4}
\usepackage{amsmath}
\usepackage{amssymb}
\usepackage{graphicx}


\begin{document}

\title{Magnon transport through a quantum dot: Conversion to electronic spin and charge currents}

\author{\L{}ukasz Karwacki}
\email{karwacki@amu.edu.pl}
\affiliation{Faculty of Physics, Adam Mickiewicz University, 61-614 Pozna\'{n}
Poland}
\author{Piotr Trocha}
\affiliation{Faculty of Physics, Adam Mickiewicz University, 61-614 Pozna\'{n}
Poland}
\author{J\'{o}zef Barna\'{s}}
\affiliation{Faculty of Physics, Adam Mickiewicz University, 61-614 Pozna\'{n}
Poland}%
\affiliation{Institute of Molecular Physics, Polish Academy of Sciences,
60-179 Pozna\'{n}, Poland}

\begin{abstract}
We consider a single-level quantum dot coupled to magnetic insulators (magnonic reservoirs) and magnetic metals (electronic reservoirs). The whole system is in an external magnetic field. In a general case, the system includes two magnonic and two electronic reservoirs, but we also present results for some specific situations, where only two or three reservoirs are effectively connected to the dot. The main objective is the analysis of the conversion of magnon current to electronic spin and charge currents, and {\it vice versa}. We consider the limiting case of large Coulomb energy in the dot (Coulomb blockade), as well as the case when the Coulomb energy is finite and double occupancy is allowed.
\end{abstract}

\maketitle

\section{Introduction}

The main carrier of information in contemporary electronics is still the electron charge, and functionality of electronic devices relies mainly on charge transport. However,  coupling between electrons and phonons in semiconducting and metallic systems leads to heat generation, when a charge current is driven in the system by an external electric field. Since the generated heat reduces functionality of a device, it must be conducted out of  the system. Reducing the size of electronic devices, as desired by the present-day nanotechnology and nanoelectronics, results in even greater problems with the generated  heat. For instance, self-heating in silicon nanowire transistors results in higher channel operating temperature, that effectively reduces mobility of the charge carriers.\cite{Mukherjee}
In very thin nanowires, the Joule heating may even lead to breakdown of the wires.~\cite{Santini}

In the past decades many attempts were undertaken in order to reduce the generated heat. As a result, a new field in electronics has emerged, known as spintronics or spin electronics, whose main objective is to use electron spin and spin current on equal footing with electron charge and charge current.~\cite{Zutic} Further development of spin electronics resulted in spin caloritronics. Its main goal is to utilize the dissipated heat energy by converting it to electric energy, which in turn can  be used for instance to drive spin currents. Thus, spin current can be driven not only by an external voltage, but also due to a temperature gradient.~\cite{Bauer} This concept has led to the discovery of various spin counterparts of conventional thermoelectric phenomena, like spin Seebeck and spin Peltier effects.~\cite{Uchida, Uchida2, Jaworski, Flipse, Xiao}

It turned out that spin waves (referred to also as magnons) can be used as carriers of physical information.
Since spin waves carry no charge, but only energy and angular momentum,\cite{Mattis} many problems with  excessive heat generation can be avoided in magnon-based devices. Apart from this, coupling of magnons to phonons is generally weaker than coupling of electrons to phonons, which may lead to different magnon and electron (phonon) temperatures.~\cite{Xiao,Agrawal}
Moreover, spin waves can propagate over relatively long distances without being scattered.~\cite{Serga}
Recently, many well-known device concepts, such as multiplexers,~\cite{Vogt} transistors,~\cite{Chumak} diodes,~\cite{Borlenghi,Borlenghi2,Ren} and  logic devices~\cite{Lenk} have been modified to exploit magnons in their working principle. However, despite the recent development and progress in this field (called also magnonics), the road to real applications of spin waves, especially in the information technology, seems to be still long. A more possible scenario is integration of electron-based and magnon-based elements into existing electronic architectures.
However, to make such an integration effective, one needs methods of converting spin waves to a spin current of electronic type, and {\it vice versa}. It has been shown that spin waves generated at a metal-insulator interface can be electrically detected in the metallic film due to the inverse spin Hall effect.~\cite{Kajiwara} This has been also shown in a spin-valve system.~\cite{Cahaya} Thus, spin waves generate a spin current in the metallic part of the system, which is then converted to a voltage signal due to the inverse  spin Hall effect. Similarly, an  electric current passing through a metallic film can excite spin-waves in the adjacent insulating magnetic layer.~\cite{Zhang} This appears when electric current generates spin current due to the spin Hall effect, and the spin current is absorbed by a magnetic layer exciting spin waves.

The magnon spin current can be manipulated by external magnetic field and also depends on such material properties like magnetic anisotropy,~\cite{Chotorlishvili} for instance. It has been also shown that the above described interface effects can lead, for instance, to rectification of thermally generated spin current.~\cite{Ren2} The aforementioned conversion of spin current requires direct contact of both layers, as the inclusion of a nonmagnetic and insulating interlayer diminishes the conversion efficiency.~\cite{Uchida2} However, conversion of spin-waves to electronic current and {\it vice versa} can be also achieved by coupling the layers through a molecule or through a quantum dot.

Quantum dots are very often utilized for investigation of quantum phenomena, mainly because their basic parameters can be easily controlled, for instance by tuning gate voltages.  As for their electronic applicability, quantum dot systems allow for single-electron transport, which ameliorates the waste heat problem more efficiently than in bulk systems. Recently, it has been shown that quantum dots can serve as efficient power generators.~\cite{Esposito,Esposito2, SanchezSothmann, SanchezSothmann2, BergenfeldtSothmann, Sothmann3} Another interesting possibility for efficient generation of charge and spin currents has been presented in quantum dot systems, where electrons interact with  bosonic particles like phonons~\cite{Wohlman} or magnons,~\cite{Sothmann,Sothmann2} and in magnonic quantum dots with magnon  transport only.~\cite{Strelcyk,Wang} More specifically, in Ref.~\onlinecite{Sothmann} a three terminal system (quantum dot in the limit of large Coulomb interaction, connected to two metallic and one insulating magnetic leads) has been studied theoretically. It has been shown that due to magnon-assisted charge transfer processes, the temperature difference between magnetic and electronic reservoirs can lead to a pure spin current and also to a spin-polarized charge current in the metallic leads. It has been also shown that in the limit of full spin polarization of the metallic leads, the efficiency of the heat to work conversion can achieve the Carnot limit in the antiparallel magnetic configuration. In Ref.~\onlinecite{Sothmann2}, in turn, charge and spin transport through a quantum dot coupled to two ferromagnetic leads with electron-magnon interaction has been considered theoretically. It has been shown, that the magnon assisted tunneling processes lead to some features in  the differential conductance. Additionally, the transport characteristics depend then on the spin  polarization of the electronic leads, leading to some asymmetrical behavior for large spin polarization and to a negative differential conductance in the parallel configuration. However, no thermal transport was considered in Ref.~\onlinecite{Sothmann2}.

In this paper we consider conversion of a spin current due to magnons to a  spin  current due to conduction electrons in a four-terminal system based on a quantum dot.
The system is presented schematically in Fig.~\ref{fig:model} and consists of a single-level quantum dot (QD) coupled to two metallic reservoirs of spin-polarized (in a general case) electrons, and to two magnetic insulators playing the role of magnon reservoirs.
Generally, adding the fourth terminal (second magnonic reservoir) provides some additional possibilities of external control of spin, charge and magnonic currents by tuning the temperature difference between the two magnonic reservoirs and also by tuning asymmetry in the coupling of these reservoirs to the dot, which may lead for instance to magnon/spin diode effects. To show this  we employ the Pauli's master equation method, which along with the model is described in detail in Sec. II. In Sec. III we present numerical results on spin current in systems of reduced geometry, i.e. in two-terminal devices. This section describes basic tunneling processes that are also present in more complex geometries investigated in the following sections. Specifically, we consider two distinct cases, where the quantum dot is coupled either to two magnonic  or to one electronic and one magnonic reservoirs. In the former case, we show that the magnonic current depends remarkably on the magnetic field applied to the system, as well as on the temperature difference between the magnonic reservoirs and  asymmetry in the dot-lead coupling. The results indicate that a quantum dot can transfer pure spin current between insulating leads and it can serve as a spin current rectifier. In turn, the system with one electronic and one magnonic leads is shown to be a simple device for conversion of magnonic to electronic spin current and {\it vice versa}. The spin current depends on the temperature difference between the two leads and also on the spin polarization of the metallic lead and on the magnetic field applied to the system. The general four-terminal case is considered in Sec. IV, where we analyze the limit of large (infinite) $U$ as well as the case of finite $U$. We focus there on the magnon to electronic spin current conversion, and analyze the dependence of spin current  on the spin polarization and magnetic configuration of the metallic leads. Apart from this, we show there how the spin current depends on the difference in temperatures of  various reservoirs and on voltage applied to the electronic ones.
In Sec. V we consider conversion of magnon current to charge current along with the influence of magnetic field on the power and efficiency of the corresponding heat engine. Summary and final conclusions are in Sec. VI.

\section{Theoretical description}
\subsection{Model}

\begin{figure}
\includegraphics[width=0.85\columnwidth]{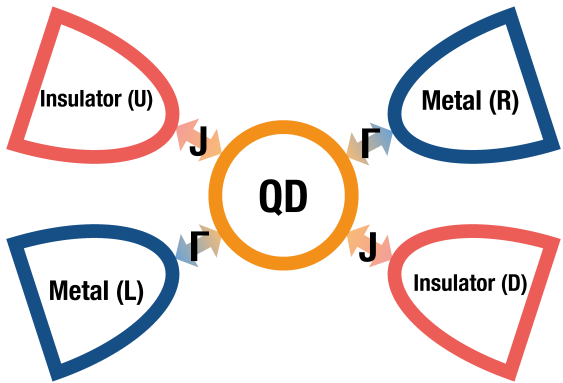}
\caption{Schematic representation of the system considered in this paper. A single-level quantum dot (QD) is coupled to two metallic leads (L,R) and to two magnetic insulators (U,D). The metallic and insulating leads are also referred to as electronic and magnonic reservoirs, respectively.  The corresponding coupling strengths are $\Gamma$ and $J$.}
\label{fig:model}
\end{figure}

The system studied in this paper is presented schematically in
Fig.~\ref{fig:model}. It is based on  a single-level quantum dot which is coupled to two ferromagnetic
metallic electrodes (reservoirs of spin polarized electrons) and to two insulating magnetic contacts (reservoirs of magnons).
The reservoirs of electrons and magnons will be referred to in the following also as electronic and magnonic reservoirs, respectively. We will consider only collinear configurations of the magnetic moments of external reservoirs. One of the best materials for the magnonic reservoirs might be YIG (yttrium-iron garnet) ferrimagnet due to its low magnetic damping and thus long
spin-wave life-time.~\cite{Serga}

The whole system under consideration can be described by a general Hamiltonian of the form
\begin{equation}\label{eq:hamiltonian}
H=H_{\rm el}+H_{\rm QD}+H^{\rm t}_{\rm el}+H_{\rm mag}+H^{\rm t}_{\rm mag},
\end{equation}
where the first term, $H_{\rm el}=\sum_{\beta\mathbf{k}\sigma}\varepsilon_{\beta\mathbf{k}\sigma}c_{\beta\mathbf{k}\sigma}^{\dagger}c_{\beta\mathbf{k}\sigma}$,
describes spin-polarized electrons in the left ($\beta=L$) and right ($\beta=R$) metallic leads. Here, $\varepsilon_{\beta\mathbf{k}\sigma}$ is the energy of electrons with wavevector $\mathbf{k}$ and spin $\sigma=\uparrow,\downarrow$ in the $\beta$-th electrode, including also the electrostatic energy shift due to a voltage applied to the system. Since the Stoner splitting in ferromagnetic metals is much larger than the Zeemann splitting due to external field, the latter may be ignored, though it may be included if necessary.

The second term in Hamiltonian (\ref{eq:hamiltonian}) describes the quantum dot and is assumed in the Anderson-type form,
\begin{equation}\label{eq:qd}
H_{\rm QD}=\sum_{\sigma}\varepsilon_{d\sigma}d_{\sigma}^{\dagger}d_{\sigma}+Un_{\uparrow}n_{\downarrow},
\end{equation}
where $\varepsilon_{d\sigma}=\varepsilon_{d}-\hat{\sigma}g\mu_BB/2$ is the
dot's level energy, whose degeneracy is lifted by an external magnetic
field $B$ ($\varepsilon_{d}$ is the bare dot's level energy). Here, $g$ is the Lande factor for the dot, $\mu_B$ is the Bohr magneton,
while $\hat{\sigma}=+(-)$ for $\sigma=\uparrow(\downarrow)$. Note,  the positive magnetic field $B$ is opposite to the axis $z$.
The second term in $H_{\rm QD}$ describes the intradot Coulomb interaction
between electrons of opposite spins.

Tunneling between the electronic reservoirs
and the dot is described by the following Hamiltonian:
\begin{equation}\label{eq:tunneling}
H^{\rm t}_{\rm el}=\sum_{\beta\mathbf{k}\sigma}V_{\beta\mathbf{k}\sigma}c_{\beta\mathbf{k}\sigma}^{\dagger}d_{\sigma}+ \mathbf{\rm{H.c.}},
\end{equation}
where $V_{\beta\mathbf{k}\sigma}$ are the corresponding tunneling matrix elements.

The term $H_{\rm mag}$ in Hamiltonian (\ref{eq:hamiltonian}) describes the top ($\alpha=U$) and bottom ($\alpha=D$) insulating magnetic components of the system, which serve as magnonic reservoirs. These components are described by the  Heisenberg model, \begin{equation}\label{eq:heisenberg}
H_{\rm mag}=\frac{1}{2}\sum_{\alpha,i, \delta}J_{\rm ex}^\alpha\mathbf{S}_{\alpha i}\cdot\mathbf{S}_{\alpha i+\delta} -g_m^\alpha\mu_BB\sum_{\alpha,i}S_{\alpha,i}^z,
\end{equation}
where summation over $\delta$ denotes summation over all nearest neighbors of a site $i$, $J_{\rm ex}^\alpha$ ($J_{\rm ex}^\alpha<0$) is the corresponding nearest-neighbour exchange integral, while $g_m^\alpha$ is the Lande factor for the $\alpha$-th magnonic reservoir. In the following we assume $J_{\rm ex}^U=J_{\rm ex}^D=J_{\rm ex}$ and $g_m^U=g_m^D=g_m$.
Difference between the Lande factors of the dot, $g$, and of the magnonic reservoirs, $g_m$, is essential from the point of view of magnon filtering, and to have a nonzero magnon current in the system considered one needs $g\ge g_m$. In the following we assume this condition to be fulfilled.~\cite{Pryor, Nilsson, Wohlfahrt}
Note, the external magnetic field is assumed the same for the quantum dot and magnonic reservoirs, and we have omitted a magnetic anisotropy (the latter can be easily included if necessary).

The last term in Hamiltonian (\ref{eq:hamiltonian}) describes exchange coupling between the quantum dot and the magnetic insulator  components,
\begin{equation}\label{eq:heisenberg2}
H^{\rm t}_{\rm mag}=\sum_{\alpha,i}j_{\rm ex}^{\alpha i}\mathbf{S}_{\alpha i}\cdot \mathbf{s},
\end{equation}
where $\mathbf{s}$ is the spin of an electron in the dot's level, and the summation over $i$ is limited to interfacial lattice sites in the magnonic reservoirs. The corresponding exchange coupling parameters are denoted as $j_{\rm ex}^{\alpha i}$. These coupling parameters for the $\alpha =U$ and $\alpha =D$  magnonic reservoirs may be different, even if the reservoirs are the same.

Performing the Holstein-Primakoff transformation,~\cite{HolsteinPrimakoff}
one may write the term $H_{\rm mag}$ as
\begin{equation}
H_{\rm mag}=\sum_{\alpha\mathbf{q}}\epsilon_{\mathbf{q}}a_{\alpha\mathbf{q}}^{\dagger}a_{\alpha\mathbf{q}},
\end{equation}
where $\epsilon_{\mathbf{q}}$  is the spin wave energy (equal in both reservoirs) for the wavevector $\bf q$, which is given by the formula (see eg. Ref.\onlinecite{Mattis}) $\epsilon_{\mathbf{q}} =2SJ\sum_{\mathbf{\delta}}[1-\cos(\mathbf{q}\cdot\mathbf{r}_\delta)] + g_m\mu_BB$, where
$\mathbf{r}_\delta$ are vectors to nearest neighbours. In turn, $H^{\rm t}_{\rm mag}$ can be written as
\begin{equation}
H^{\rm t}_{\rm mag} = \sum_{\alpha\mathbf{q}}j_{\alpha\mathbf{q}}a_{\alpha\mathbf{q}}^{\dagger}d_{\uparrow}^{\dagger}d_{\downarrow}+\mathbf{ \rm{H.c.}},
\end{equation}
where $j_{\alpha\mathbf{q}}$ depends generally on the distribution of interfacial spins and also on coupling between these spins and the quantum dot. The explicit form of $j_{\alpha\mathbf{q}}$ is not presented here as this coupling will be treated as a parameter (see below).

\subsection{Method}

In order to calculate the charge, spin and magnon currents one needs to find first the probabilities $P_{i}$ for all available dot's states $|i\rangle$.
To do this we use the Pauli's master equation method, based on the weak coupling and Markov approximations.~\cite{Flensberg,Timm}
The master equation takes the form
$\dot{P}_{i}=\sum_{j}\left(W_{ji}P_{j}-W_{ij}P_{i}\right)$,
where $W_{ij}$ is the transition rate from the dot's state $|i\rangle$ to the state $|j\rangle$. This transition rate is given by the Fermi golden rule as,
\begin{equation}\label{eq:rates}
W_{ij}=\frac{2\pi}{\hbar}\sum_{mn}|\langle n;j|H|i;m\rangle|^{2}w_{m}\,\delta\left(E_{m,i}-E_{n,j}\right),
\end{equation}
and is a sum  of partial transition rates from an initial many body state $|i;m\rangle$ with energy $E_{i,m}$ to a state $|j;n\rangle$ with energy $E_{j,n}$. The state $|i;m\rangle$ is a state of the whole system, and indicates that the dot is in the state $|i\rangle$, while the electrodes are in the state $|m\rangle$. Furthermore,
$w_{m}$ denotes the probability of finding the electrodes  in the initial state $|m\rangle$.

The master equation  can be written in a matrix form as
$\dot{\mathbf{P}}=\tilde{\mathbf{W}}\mathbf{P}$,
which in the stationary state considered here simplifies to
\begin{equation}\label{eq:mastereq}
\tilde{\mathbf{W}}\mathbf{P}=\mathbf{0}.
\end{equation}
The matrix $\tilde{\mathbf{W}}$ is determined by the transition rates (8). Additionally, the probabilities $P_i$ obey the normalization condition, $\sum_{i}P_{i}=1$.

To determine the transition rates from Eq.~(\ref{eq:rates}), we employ the wide-band approximation, which is based on the assumption that the  coupling of the dot's and leads' states is independent of energy in the electron or magnon bands. This allows us to write the dot level widths $\Gamma_{\beta\sigma}$ due to coupling to the electronic reservoir $\beta$ as
$\Gamma_{\beta\sigma} = 2\pi\langle|V_{\beta{\mathbf{k}}\sigma}|^{2}\rangle\rho_{\beta\sigma}=(1\pm p_{\beta})\Gamma_{\beta}$. These parameters will be considered as effective coupling parameters to the electronic reservoirs. Similarly, the excited state in the dot has also a finite life time due to coupling to the magnonic reservoirs, and the corresponding contribution to the level width can be written as
 $J_{\alpha} = 2\pi\langle|j_{\alpha\mathbf{q}}|^{2}\rangle\rho_{\alpha}$, which will be treated as effective coupling parameters between the dot and magnonic reservoirs.
 Above, $\langle|V_{\beta\mathbf{k}\sigma}|^{2}\rangle$ and $\langle|j_{\alpha\mathbf{q}}|^{2}\rangle$ are the corresponding averages over ${\bf k}$ and ${\bf q}$, respectively. Apart from this, $\rho_{\beta\sigma}$ and $p_{\beta}$ stand for the density of electron states and spin polarization in the metallic lead  $\beta$, while $\rho_{\alpha}$ is the density of magnon states in the magnetic insulating leads.

In order to determine the matrix $\tilde{\mathbf{W}}$ in Eq.~(\ref{eq:rates}), we need to know the occupation of electron states in the electronic reservoirs, $\langle c_{\beta\mathbf{k}\sigma}^{\dagger}c_{\beta\mathbf{k}\sigma} \rangle = f_{\beta\sigma}^{+}(\varepsilon_{\beta\mathbf{k}\sigma})$, which is given by
the Fermi-Dirac distribution function
$f_{\beta\sigma}^{+}(\varepsilon)  =  1/[\exp(\frac{\varepsilon-\mu_{\beta\sigma}}{k_BT_{\beta}})+1]\equiv 1-f_{\beta\sigma}^{-}(\varepsilon)$,
where $\mu_{\beta\sigma}$ is the electrochemical potential in the electrode $\beta$ for spin $\sigma$, while $T_\beta$ is the corresponding temperature (equal in both spin channels).
The spin dependence of chemical potential may result from externally applied spin bias or from spin accumulation effects. In the following, however, the spin accumulation will be omitted. The conventional voltage is then $V=(\mu^0_L-\mu^0_R)/e$, while the spin voltage is equal to $V^s=(\mu^s_L-\mu^s_R)/e$, where $\mu^0_\beta=(\mu_{\beta \uparrow}+
\mu_{\beta \downarrow})/2$ and $\mu^s_\beta=(\mu_{\beta \uparrow}-
\mu_{\beta \downarrow})/2$ for $\beta =L,R$,  while $e$ is electron charge, $e<0$.
In turn, population of magnons in the magnonic reservoirs, $\langle a_{\alpha\mathbf{q}}^{\dagger}a_{\alpha\mathbf{q}} \rangle\equiv n_{\alpha}^{+}(\epsilon_{\alpha{\bf q}})$, is determined by the Bose-Einstein distribution function
$n_{\alpha}^{+}(\epsilon)  =  1/[\exp(\frac{\epsilon}{k_BT_{\alpha}})-1]\equiv n_{\alpha}^{-}(\epsilon)-1$.

Generally, the temperatures and electrochemical potentials of the electronic reservoirs are different. We neglect in the following the spin voltage, so one can write
\begin{subequations}
 \begin{align}
\mu_{\beta\sigma} & =  \mu^{0}\mp\Delta\mu/2\,, \\
T_{\beta} & = T_{\rm el}^{0}\pm\Delta T_{\rm el}/2\,,
 \end{align}
\end{subequations}
where the upper and lower signs correspond to the left ($\beta =L$) and right ($\beta=R$) reservoir, respectively.
In turn, for the magnonic reservoirs we write
\begin{equation}
T_{\alpha}  = T_{\rm mag}^{0}\pm\Delta T_{\rm mag}/2,
\end{equation}
where the upper and lower signs correspond to the upper ($\alpha =U$) and bottom ($\alpha =D$) magnonic reservoir, respectively.
Accordingly, we can write
$\Delta T_{\rm el} = T_{L}-T_{R}$ and $\Delta T_{\rm mag}  = T_{U}-T_{D}$ for the difference in temperatures of the electronic and magnonic reservoirs, and
$\Delta\mu =  \mu_{R}-\mu_{L}=eV$  for the difference in electrochemical potentials of the electronic reservoirs, where $V$ is the externally applied voltage
(positive for current flowing from the left to right reservoir).

\subsection{Charge and spin currents}

Charge current flowing to the dot is a sum of currents flowing from the left, $I^L$, and from the right, $I^R$, electronic reservoirs. From the charge conservation one finds
\begin{equation}
I^{L}+I^{R}=e\frac{d\langle N\rangle}{dt} = e\left(\dot{P}_{\uparrow}+\dot{P}_{\downarrow}\right),
\end{equation}
where $e\langle N\rangle=e(P_{\uparrow}+P_{\downarrow})$ is the average charge in the dot. In turn, the current flowing from the reservoir $\beta$ is the sum of currents flowing in the spin-up and spin-down channels,
$I^{\beta}=I_{\uparrow}^{\beta}+I_{\downarrow}^{\beta}$.
In the stationary state, the above equation gives
\begin{eqnarray}\label{eq:conservation}
I^{L}=-I^{R}.
\end{eqnarray}
Thus, the formula for charge current flowing from the left to right electron reservoirs can be further symmetrized as $I=(I^{L}-I^{R})/2$.
What we need now is the formula for $I_{\sigma}^{\beta}$, which can be found taking into account the relevant transition rates.

In the following we will consider separately the situation when the current involves only empty and singly occupied dot states (due to Coulomb blockade), and the situation when  double occupancy is also allowed. Thus, we split the general formula for current into two terms. The first term, $I_{\sigma}^{\beta(1)}$, describes contribution of tunneling processes through zero and singly occupied dot, while the second term, $I_{\sigma}^{\beta(2)}$, represents contribution from tunneling processes involving  double occupancy. Thus, we write
$I_{\sigma}^{\beta}=I_{\sigma}^{\beta(1)}+I_{\sigma}^{\beta(2)}$.
Taking into account the expressions for transition rates, the first term can be written in the form
\begin{equation}
I_{\sigma}^{\beta(1)}=e\left(P_{0}W_{0\sigma}^{\beta}-P_{\sigma}W_{\sigma0}^{\beta}\right),
\end{equation}
while the second contribution takes the form
\begin{equation}
I_{\sigma}^{\beta(2)}=e\left(P_{\overline{\sigma}}W_{\overline{\sigma}2}^{\beta}-P_{2}W_{2\overline{\sigma}}^{\beta}\right).
\end{equation}

The total spin current $J_{\rm s}$ flowing to the quantum dot includes the spin currents  $J_{\rm s}^L$ and $J_{\rm s}^R$ flowing from the left and right electronic reservoirs,
as well as the spin currents flowing from the top and bottom  magnonic ones, $J_{\rm s}^U$ and $J_{\rm s}^D$. Thus, one can write
$J_{\rm s}=J_{\rm s}^L+J_{\rm s}^R + J_{\rm s}^U+J_{s}^D$. From the spin angular momentum conservation
\begin{eqnarray}
J_{\rm s}^L+J_{\rm s}^R + J_{\rm s}^U+J_{\rm s}^D =\frac{d\langle s_{z}\rangle}{dt}&=& \frac{\hbar}{2}\left(\dot{P}_{\uparrow}-\dot{P}_{\downarrow}\right) ,
\end{eqnarray}
where $\langle s_{z}\rangle=(\hbar /2)(P_{\uparrow}-P_{\downarrow})$ is the average spin momentum in the dot.
In the  stationary state one finds,
\begin{equation}
J_{\rm s}^{L}+J_{\rm s}^{R}=-(J_{\rm s}^{U}+J_{\rm s}^{D}).
\end{equation}

The spin currents flowing from electronic reservoirs are determined by the corresponding spin-polarized charge currents,
\begin{equation}
J_{\rm s}^{\beta}= \frac{\hbar}{2e}(I_{\uparrow}^{\beta}-I_{\downarrow}^{\beta})
\end{equation}
for $\beta = L,R$,
while the magnonic contributions can be calculated from the formula
\begin{equation}
J_{\rm s}^{\alpha} = -\hbar \left(P_{\uparrow}W_{\uparrow\downarrow}^{\alpha}-P_{\downarrow}W_{\downarrow\uparrow}^{\alpha}\right)
\end{equation}
for $\alpha =U,D$. Note, the spin current from magnonic reservoirs can flow only when the dot is occupied by a single electron, while no spin current can flow when the dot is either empty or doubly occupied.

One can also calculate heat fluxes associated with electronic and magnon currents.
To do this we take into account the thermodynamic law,
$\langle\dot{Q}\rangle = \langle\dot{E} \rangle - \langle\dot{W}\rangle$,
where $\langle\dot{E} \rangle$ is the rate of internal energy increase while $\langle\dot{Q}\rangle$ and $\langle\dot{W} \rangle$ are the heat delivered to the system and work done on it in a unit time. Thus, in the stationary state the heat current flowing out of the lead $\beta (=L,R)$, associated with electronic current,
can be written in the form
\begin{equation}
J_Q^{\beta}=\frac{1}{e}\sum_\sigma\left[ (\varepsilon_{\sigma}-\mu_{\beta\sigma})I_{\sigma}^{\beta(1)}+(\varepsilon_{\sigma}+U-\mu_{\beta\sigma})I_{\sigma}^{\beta(2)}\right] .
\end{equation}
In turn, the heat current flowing from magnonic reservoir $\alpha (=U,D)$ can be written as
\begin{equation}
J_Q^{\alpha}=g\mu_B B\,I_{\textrm{mag}}^{\alpha}.
\end{equation}

Below we present some numerical results. First, in Sec. III we present results on spin current in a reduced two-terminal geometry. Then in Sec. IV we will consider the general four-terminal case.

\section{Results: two-terminal geometry}

In this section we present numerical results on spin transport in two two-terminal systems. Since magnons can flow only when the dot is singly occupied, we assume in this section the limit of large Hubbard parameter, $U\to \infty$, i.e., when the dot can be occupied at most by a single electron. This section is divided into two parts, each of them corresponding to a specific configuration of the dot and electronic and magnonic reservoirs. In the first part we consider pure magnon transport through a quantum dot coupled to two magnonic reservoirs (I-QD-I system), i.e. the electronic reservoirs are decoupled from the dot. We focus there on the influence of magnetic field $B$ and  difference  $\Delta T_{\textrm{mag}}$ in temperatures of the two magnonic reservoirs on the magnon current.
Then, one magnonic reservoir is replaced by an electronic reservoir, and the system is referred to as the M-QD-I system. Our main interest is in conversion of magnonic spin current into electronic spin current, and {\it vice versa}. Here, one of the key parameters modifying the spin current is the polarization parameter $p$ of the metallic lead. Obviously, in both cases there is no charge current.
Let us  start from the system, where the dot is attached to two magnonic reservoirs only.

\subsection{Case of $\Gamma_{L}=\Gamma_{R}=0$: {\rm I-QD-I} system}

\begin{figure*}
\includegraphics[width=.89\textwidth]{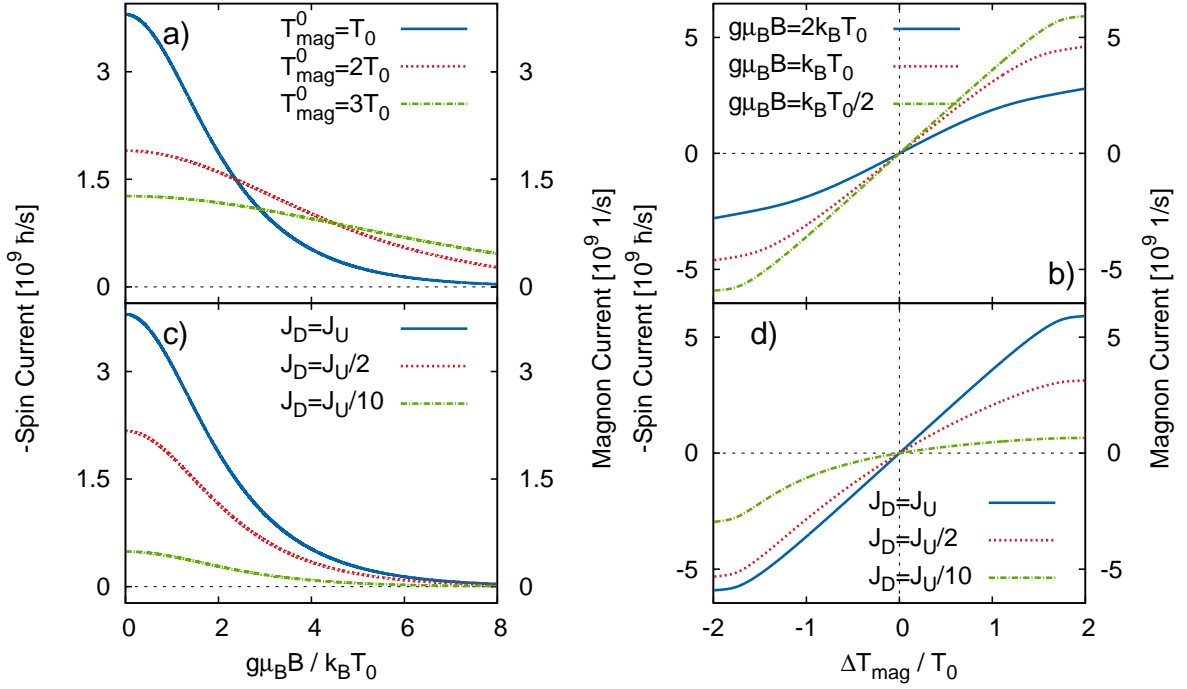}
\caption{I-QD-I case. Spin current (and the corresponding magnon current)  as a function of magnetic field $B$ (a) and (c), and as a function of the  difference $\Delta T_{\rm mag}$ in temperatures of the magnonic reservoirs (b) and (d). The other parameters are:
(a) $J_{D}=J_{U}=0.1k_BT_{0}$, $\Delta T_{\rm mag}=T_{0}$; (b) $J_{D}=J_{U}=0.1k_BT_{0}$, $T_{\rm mag}^{0}=T_{0}$; (c)
$T_{\rm mag}^{0}=T_{0}$, $\Delta T_{\rm mag}=T_{0}$; and (d) $g\mu_{B}B=k_BT_{0}/2$, $T_{\rm mag}^{0}=T_{0}$. For
all figures $\varepsilon_{d}=0$ and $k_BT_{0}=0.1$ meV.}
\label{fig:magnon1}
\end{figure*}

When considering the quantum dot coupled to two
magnonic reservoirs only, we assume that the dot is initially prepared in one
of the two spin states, $\lvert\uparrow\rangle$ or $\lvert\downarrow\rangle$. Such a single-electron state is required
in order to mediate the magnon transport between the two magnonic reservoirs.
Transport of magnons through the dot does not change its charge state, so the dot remains singly occupied
and only its  spin state varies due to the magnon current. The corresponding stationary occupation
probabilities, $P_{\uparrow}$ and $P_{\downarrow}$, can be found from the master equation (\ref{eq:mastereq}).
To solve this equation we need the matrix $\tilde{\mathbf{W}}$, which in the case under consideration acquires the form
\begin{equation}
\tilde{\mathbf{W}}=\frac{1}{\hbar}\sum_{\alpha}
\left[ \begin{array}{cc}
-J_{\alpha}n_{\alpha}^{+} & J_{\alpha}n_{\alpha}^{-} \\
J_{\alpha}n_{\alpha}^{+} & -J_{\alpha}n_{\alpha}^{-}
\end{array}\right].
\end{equation}

The spin current conservation takes now the form $J_{\rm s}^{U}=-J_{\rm s}^{D}$. This allows for symmetrization of the formula for spin current flowing from the top ($U$) magnonic reservoir to the  bottom ($D$) one,  $J_{\rm s}^{U\to D}=(J_{\rm s}^{U}-J_{\rm s}^{D})/2$. Accordingly, the final expression for the spin current associated with a flow of magnons from the top ($U$) to the  bottom ($D$) magnonic reservoir can be written as follows:
\begin{equation}
J_{\rm s}^{U\to D}=J_{D}J_{U}\frac{n_{D}^{+}-n_{U}^{+}}{J_{D}(1+2n_{D}^{+})+J_{U}(1+2n_{U}^{+})},
\end{equation}
where $n_{U}^{+}$ and $n_D^{+}$ are the Bose-Einstein distributions of magnons in the top and bottom magnonic reservoirs, respectively,
which depend on the Zeeman splitting of the dot's energy level,  $n_{U}^{+}=n^+_{U}(\epsilon=g\mu_BB)$ and $n^{+}_{D}=n^+_{D}(\epsilon=g\mu_BB)$.

There are three conditions for the magnon transport to occur in this configuration. First, the quantum dot energy level has to be non-degenerate, which is achieved by the Zeeman splitting due to an external magnetic field. Second, one needs magnons with energy equal to the Zeeman splitting of the dot level, which is assured by the condition $g_m\le g$. The magnetic field effectively serves then as a magnon energy filter. Third, the magnons can be transported from one reservoir to the other when there is an imbalance in the magnon distributions between the top and bottom  magnonic reservoirs. This imbalance appears when there is a difference in temperatures of the reservoirs. The magnons are then transported from the reservoir of higher temperature to the reservoir of lower temperature. Individual magnon transport processes are accompanied by spin-flip processes in the dot.
Since each magnon caries the spin angular momentum with the $z$-component equal to $-\hbar$ (we consider the case of $B>0$), the magnon current $j_{\rm mag}$, defined as the number of magnons transmitted from the top to the bottom reservoir  in a unit time, is equal to the corresponding spin current divided by $-\hbar$, i.e., $j^{U\to D}_{\rm mag}=J_{\rm s}^{U\to D}/(-\hbar)$.
Thus, in the situation under consideration the spin current and magnon current have opposite signs.

Figure~\ref{fig:magnon1}(a) presents the magnon current $j^{U\to D}_{\rm mag}$ and also the associated spin current $J_{\rm s}^{U\to D}$ as a function of magnetic field $B$ for indicated values of the average temperature $T_{\rm mag}^{0}$, and for $\Delta T_{\rm mag}=T_{0}$. Since the spin current in this figure is measured in the units of  $\hbar/s$, the relevant curves show simultaneously both spin and magnon currents (but they differ in sign). Therefore, when describing numerical results we preferably refer to magnon current.
For $B=0$ the current is not well defined due to zero magnon energy and the associated divergency in the Bose-Einstein distribution function. Note, that only magnons of energy equal to the Zeeman splitting of the dot's level can be  transported through the dot. The magnon current in the limit of $B=0$ can be defined as
\begin{align}
j^{U\to D}_{\rm mag}(B=0) &= \lim_{B\to 0}j^{U\to D}_{\rm mag}(B) \nonumber \\ & = \frac{\frac{\Delta T_{\rm mag}}{T_{\rm mag}^{0}}J_{D}J_{U}}{\frac{\Delta T_{\rm mag}}{T_{\rm mag}^{0}}\left(J_{U}-J_{D} \right)+2\left(J_{U}+J_{D} \right)}.
\end{align}

Note, that such a  problem will be absent when the magnetic anisotropy is included, so the  magnon energy is nonzero for $B=0$, independently of the wavevector. As shown in Fig.~\ref{fig:magnon1}(a),  the magnon current achieves a maximum in the limit of zero field, $B=0$, and disappears for large values of $B$. This behavior follows from the Bose-Einstein distribution function,
which leads to a high density of low-energy magnons and small density of magnons with high energy. Thus, an increase in magnetic field leads to transmission  of the magnons with higher energy, which  results in a decrease in the magnon current with increasing magnetic field $B$. Note, the magnon energy also increases with the magnetic field. Different curves in Fig.~\ref{fig:magnon1}(a) correspond to different values of $T_{\rm mag}^{0}$.  When $T_{\rm mag}^{0}$, increases, the maximum current at $B=0$ decreases, and this decrease is associated with an increasing role of magnons flowing in the opposite direction, i.e., from the bottom reservoir ($D$) to the top ($U$) one. %Note, $\Delta T_{\rm mag}$ is constant, so $\Delta T_{\rm mag} /T^0_{\rm mag}$ decrease with increasing $T_{\rm mag}^{0}$.

Figure~\ref{fig:magnon1}a corresponds to the situation when the quantum dot is symmetrically coupled to the reservoirs, $J_U=J_D$. When this coupling decreases, the magnon current also decreases. In Fig.~\ref{fig:magnon1}(c) we show the magnon current, when coupling to one of the magnonic reservoirs becomes reduced, while coupling to the other one is constant. As one might expect, variation of  the current with increasing $B$ is similar for all values of the coupling parameter, except that the maximum value of the current at $B=0$ becomes reduced when the coupling decreases.

The magnon current depends on the difference in temperature $\Delta T_{\rm mag}$ of the magnonic reservoirs. This dependence is shown in Fig.~\ref{fig:magnon1}(b) for equal couplings to the reservoirs and for indicated values of the magnetic field. The magnon current increases roughly linearly with $\Delta T_{\rm mag}$ at small values of $\Delta T_{\rm mag}$, and then the rate of increase becomes smaller at higher values of the temperature difference. This nonlinear increase is a consequence of the magnon distribution. Of course, the magnon current changes sign when the temperature difference is reversed. For all values of the magnetic field assumed in Fig.~\ref{fig:magnon1}(b), the current is symmetric with respect to reversal of the temperature bias, i.e., the absolute magnitude of current is independent of the sign of $\Delta T_{\rm mag}$.

The situation changes qualitatively when the coupling  to the top and bottom magnonic reservoirs are different, as shown in Fig.~\ref{fig:magnon1}(d). One finds then a remarkable asymmetry in the magnon current with respect to the sign reversal of $\Delta T_{\rm mag}$.  This difference is especially large for large asymmetry in the coupling parameters. For the parameters assumed in  Fig.~\ref{fig:magnon1}(d), the absolute values of magnon current for positive temperature difference, $\Delta T_{\rm mag}>0$, is smaller than for negative $\Delta T_{\rm mag}$. This is because transport of magnons is determined mainly by the smaller coupling between the dot and the reservoir. The magnon tunneling from the corresponding reservoir is then more probable when it has higher temperature, than the magnon tunneling from the dot to this reservoir when it has lower temperature. Obviously, the asymmetry disappears for symmetrical coupling.

\subsection{Case of $\Gamma_{R}=J_{D}=0$: {\rm M-QD-I} system}

Now, we consider a quantum dot coupled to one electronic (say the one corresponding to $\beta = L$) and one magnonic (corresponding to $\alpha = U$) reservoirs. It is also convenient to adapt the notation in this part to the present situation and denote the temperature of the magnonic reservoir as $T_{\rm mag}$, temperature of the electronic reservoir as $T_{\rm el}$, and the difference in temperatures of the two reservoirs as $\Delta T=T_{\rm mag}-T_{\rm el}$, while $(T_{\rm mag}+T_{\rm el})/2=T_0$.

\begin{figure}
\includegraphics[width=0.9\columnwidth]{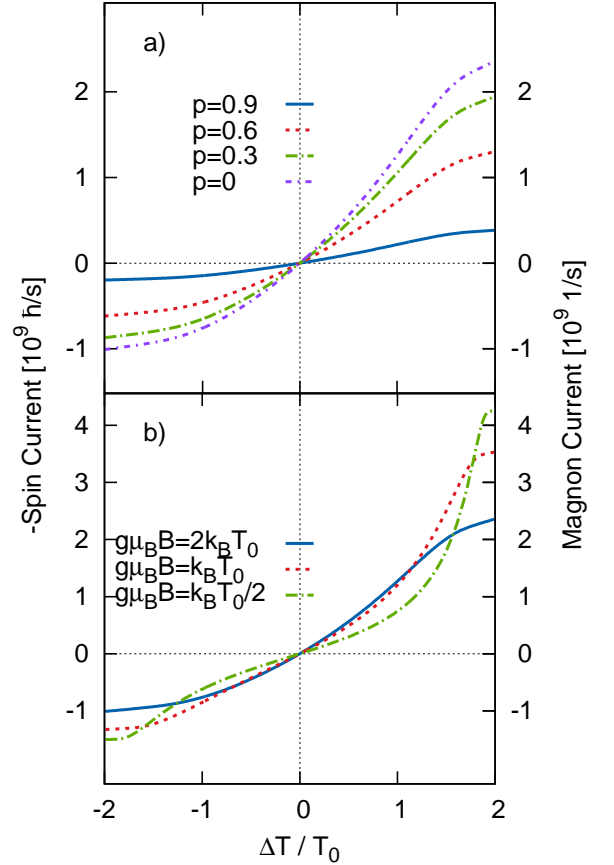}
\caption{M-QD-I system. Spin current flowing from the magnonic to the electronic reservoir (and the associated magnon current flowing from the magnonic reservoir) as a function of the normalized temperature difference $\Delta T/T_0$;  (a) for indicated values of the spin polarization $p$ of electronic reservoir and $g\mu_BB=2k_BT_{0}$; and (b) for different values of the magnetic field $B$ and $p=0$. The other parameters are $\varepsilon_{d}=0$, $J_U=\Gamma_L=0.1k_BT_{0}$, and $k_BT_{0}=0.1$ meV. The quantum dot plays a role of spin current converter from electronic to magnonic origin, and {\it vice versa}.  }
\label{fig:magnon2}
\end{figure}

Similarly as in the situation considered above, there is no charge current flowing through the dot. In turn, transport of spin between the reservoirs is admitted. However, the spin current flowing from the electronic reservoir  is of electronic type, while the spin current flowing from the magnonic reservoir is of  the magnon origin.
The spin current conservation,  Eq.~(17), takes now the form $J_{s}^{U}=-J^{L}_s\equiv J_{\rm s}^{U\to L}$, where $J_{\rm s}^{U\to L}$ is the spin current flowing from the magnonic ($\alpha =U$) reservoir to the electronic ($L$) one. Thus, the spin current associated with transport of magnons is converted to a pure spin current of electronic type, and {\it vice versa}, transport of spin current of electronic type is transformed to magnon current. The spin current $J_{\rm s}^{U\to L}$ flowing from the magnonic to the electronic reservoir can be expressed by the following formula:
\begin{widetext}
\begin{equation}
J_{\rm s}^{U\to L}=\frac{J_{U}\Gamma_{L\uparrow}\Gamma_{L\downarrow}\left[f_{L\downarrow}^{+} \left(f_{L\uparrow}^{+}-1\right)+n_{U}^{+}\left(f_{L\uparrow}^{+}-f_{L\downarrow}^{+} \right) \right]}{J_{U}\Gamma_{L\downarrow}\left[f_{L\downarrow}^{+}+n_{U}^{+}\left(1+f_{L\downarrow}^{+}\right)\right]
+J_{U}\Gamma_{L\uparrow}\left[1+n_{U}^{+}\left(1+f_{L\uparrow}^{+}\right)\right]+2\Gamma_{L\uparrow}\Gamma_{L\downarrow}\left(1-
f_{L\downarrow}^{+}f_{L\uparrow}^{+}\right)},
\end{equation}
\end{widetext}
where $f_{L\uparrow}^{+} =f_{L\uparrow}^{+}(\varepsilon = \varepsilon_\uparrow )$, $f_{L\downarrow}^{+} =f_{L\downarrow}^{+}(\varepsilon = \varepsilon_\downarrow )$, and $n_{U}^{+} = n_{U}^{+}(\epsilon = g\mu_BB)$. In turn, the magnon current flowing from the magnonic reservoir to the dot, $j^{U}_{\rm mag}$,  is then related to the spin current $J_{\rm s}^{U\to L}$ {\it via} the formula $j^{U }_{\rm mag} = J_{\rm s}^{U\to L}/(-\hbar )$.

Figure~\ref{fig:magnon2}(a) shows the spin current flowing from the magnonic to electronic reservoir as a function of the difference in temperatures of the two reservoirs. Different curves correspond to indicated values of the spin polarization factor $p$ of the electronic reservoir. If both  temperatures are equal, $T_{\rm mag}=T_{\rm el}$, there is no spin current flowing through the system. However, when $\Delta T >0$, i.e., $T_{\rm mag}>T_{\rm el}$, magnons can flow from the magnonic reservoir to the dot, and then to the metallic reservoir as a pure spin current of electronic type. We recall that the spin current has sign opposite to that of the magnon current flowing from the magnonic reservoir to the dot. On the other hand, when the temperature of the electronic reservoir is higher than that of the magnonic one, $T_{\rm mag}<T_{\rm el}$, the spin-flip processes on the dot excite magnons in the magnonic reservoir. The magnon current is then negative according to our definition, while the associated spin current is positive, see Fig.~\ref{fig:magnon2}(a).

The effect of spin polarization of the electronic reservoir on the spin current is rather clear. When the electronic reservoir is half-metallic, i.e., $p=1$, the spin current is completely suppressed. This is because electron transitions between the dot and electronic reservoir are not able to create spin-flip processes required for magnon current. The absolute value of spin current increases when the polarization $p$ decreases, and reaches a maximum value in the case of a nonmagnetic electronic reservoir, $p=0$.

Filtering of low energy magnons results in a magnon current whose absolute magnitude for $|\Delta T|$ above $T_0$  increases with decreasing magnetic field, as presented in Fig.~\ref{fig:magnon2}(b). The opposite tendency can be observed for $|\Delta T|$ smaller than $T_0$. This behavior  can be explained in a similar way as in the case of two magnonic reservoirs, presented and discussed above, and  is a results of the interplay of the fermionic and bosonic distributions.

\section{Four-terminal case: spin current}

Let us consider now the general situation, i.e., the full four-terminal system consisting of two electronic and two magnonic reservoirs, as described in Sec. II and shown explicitly in Fig. 1. First, we consider conversion of  magnon current to spin current of electronic type in the limit of large $U$ as well as  in the case of finite $U$. Then, in  the next section we analyze conversion of magnon current to charge current.

The corresponding transition rate matrix $\tilde{\mathbf{W}}$ can be  written in the following form:
\begin{widetext}
\small
\begin{equation}
\tilde{\mathbf{W}}=\frac{1}{\hbar}
\left[\begin{array}{cccc}
-\sum_{\beta,\sigma}\Gamma_{\beta\sigma}f_{\beta\sigma}^{+} & \sum_{\beta}\Gamma_{\beta\uparrow}f_{\beta\uparrow}^{-} & \sum_{\beta}\Gamma_{\beta\downarrow}f_{\beta\downarrow}^{-} & 0 \\
\sum_{\beta}\Gamma_{\beta\uparrow}f_{\beta\uparrow}^{+} & -\sum_{\beta}\left(\Gamma_{\beta\uparrow}f_{\beta\uparrow}^{-}+\Gamma_{\beta\downarrow}f_{\beta\downarrow}^{U+}\right)-\sum_{\alpha}J_{\alpha}n_{\alpha}^{+} & \sum_{\alpha}J_{\alpha}n_{\alpha}^{-} & \sum_{\beta}\Gamma_{\downarrow\beta}f_{\beta\downarrow}^{U-} \\
\sum_{\beta}\Gamma_{\beta\downarrow}f_{\beta\downarrow}^{+} & \sum_{\alpha}J_{\alpha}n_{\alpha}^{+} & -\sum_{\beta}\left(\Gamma_{\beta\downarrow}f_{\beta\downarrow}^{-}+\Gamma_{\beta\uparrow}f_{\beta\uparrow}^{U+}\right)-\sum_{\alpha}J_{\alpha}n_{\alpha}^{-} & \sum_{\beta}\Gamma_{\uparrow\beta}f_{\beta\uparrow}^{U-} \\
0 & \sum_{\beta}\Gamma_{\downarrow\beta}f_{\beta\downarrow}^{U+} & \sum_{\beta}\Gamma_{\uparrow\beta}f_{\beta\uparrow}^{U+} & -\sum_{\beta,\sigma}\Gamma_{\beta\sigma}f_{\beta\sigma}^{U-} \\
\end{array}\right],
\end{equation}
\end{widetext}
where $f_{\beta\uparrow(\downarrow)}^{U\pm}$ is defined as $f_{\beta\uparrow(\downarrow)}^{U\pm}=f_{\beta\uparrow(\downarrow)}^{\pm}(\varepsilon_{\uparrow(\downarrow)}+U)$ whereas $f_{\beta\uparrow}^{\pm}$, $f_{\beta\downarrow}^{\pm}$, and $n_{\alpha}^{\pm}$ are defined as in Sec. III.
From this matrix one can find the stationary occupation probabilities of the dot states, and then charge and spin currents flowing in the system.
%Consider now spin and magnon currents generated by thermal effects.
Let us begin with the case of  large $U$, assuming that the quantum dot can be occupied
at most by a single electron (the limit of large Hubbard parameter $U$).

\subsection{Limit of large $U$}

\begin{figure*}
\includegraphics[scale=0.75]{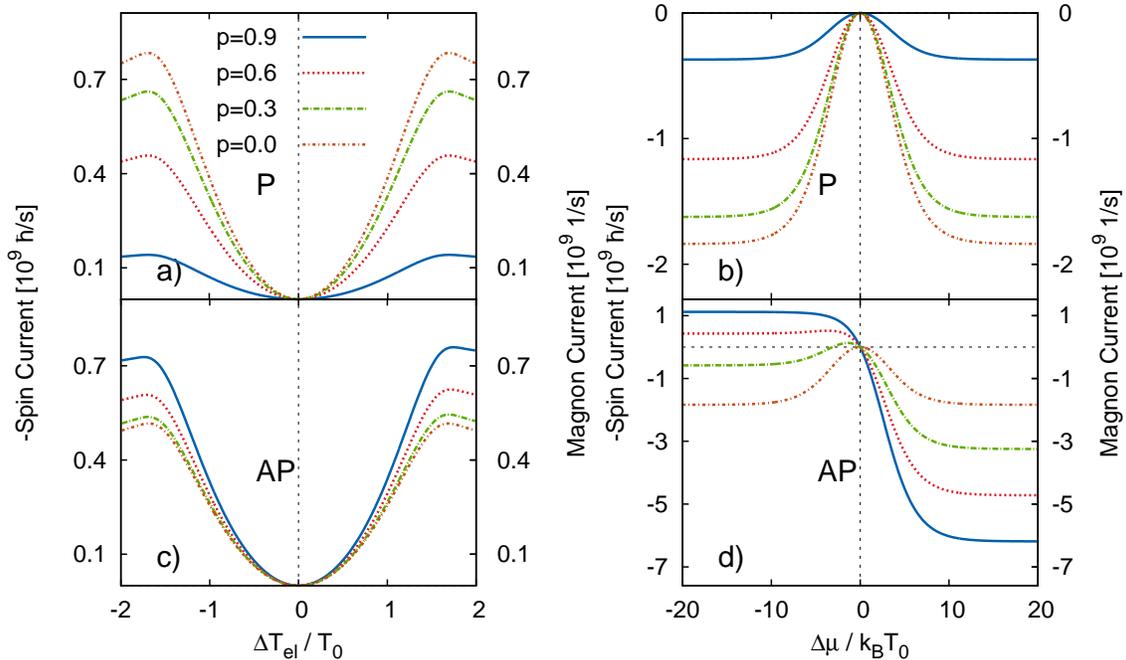}
\caption{\textit{Four-terminal case}. Total spin current $J_{\rm s}^{U}+J_{\rm s}^{D}$ (and the corresponding magnon current) flowing from the magnonic reservoirs to the electronic ones, presented as a function of the temperature difference $\Delta T_{\rm el}$ of the electronic reservoirs (a) and (c), and as a function of the difference $\Delta \mu$ in the electrochemical potentials (b) and (d),  in the parallel (P) and antiparallel (AP) magnetic configurations of the metallic leads and for indicated values of the leads' polarization $p_{L}=p_{R}=p$. The other parameters are: $\varepsilon_{d}=0$, $J_{U}=J_{D}=\Gamma_{L}=\Gamma_{R}=0.1k_BT_{0}$, $T_{\rm mag}^{0}=T_{\rm el}^{0}=T_{0}$, $\Delta T_{\rm mag}=0$, $g\mu_BB=2k_BT_{0}$, and $k_BT_{0}=0.1$ meV. Apart from this, $\Delta\mu=0$ (a) and (c), and $\Delta T_{\rm el}=0$ (b) and (d). }
\label{fig:magnon4}
\end{figure*}

 The total spin current, $J_{\rm s}^U+J_{\rm s}^D$, (and also magnon current) flowing from the magnonic reservoirs to the dot and then to the electronic reservoirs shown in Figs.~\ref{fig:magnon4}(a) and \ref{fig:magnon4}(c) for the parallel (P) and antiparallel (AP) magnetic configurations of the electronic reservoirs, respectively.  The spin (magnon) current is plotted there as a function of the difference $\Delta T_{\rm el}$ in temperatures of the electronic reservoirs, and different curves correspond to indicated values of the spin polarization factor $p$ (equal for both electronic reservoirs). We note that the AP configuration corresponds to the reversed magnetic moment of the right ($R$) electronic reservoir. Moreover, in Fig.~\ref{fig:magnon4} we assumed that both magnonic reservoirs have equal temperatures, $\Delta T_{\rm mag}=0$,  and also $T^0_{\rm mag}=T^0_{\rm el}$. Thus, when $\Delta T_{\rm el}=0$, the system is in equilibrium and neither spin nor charge currents can flow. However, an increase in $|\Delta T_{\rm el}|$ leads to the generation of positive magnon current (negative spin current). In a general case, the magnon-electron processes in the four-terminal case are more complex than in the simplified two-terminal system considered above.

For relatively large values of the spin polarization factor $p$, magnon current in the parallel (P) configuration is generally smaller than in the antiparallel (AP) one. Physical origin of this behavior is similar to that analyzed in the case of the M-QD-I system.  Tunneling probability of majority-spin electrons to the dot is then dominant. Thus, an electron that has tunneled to the dot and changed its spin orientation due to a magnon  absorption (or creation)  has  a larger probability to tunnel off the dot in the antiparallel configuration than in the parallel one.  Accordingly, the magnon current is larger in the AP configuration than in the P one.
This magnon current is converted to spin current of electronic type. Interesting situation occurs for $p=0$, when the spin current associated with magnons is the only source of spin current flowing to the electronic reservoirs. Obviously, currents in the P and AP configurations are equivalent for $p=0$ [compare Fig.~\ref{fig:magnon4}(a) and Fig.~\ref{fig:magnon4}(c) for $p=0$]. When, in turn, $p\to 1$, the magnon current in the AP configuration is nonzero, while it disappears in the P configuration. The positive sign  of the magnon current means that magnons flow to the dot, and the magnon absorption processes dominate over the magnon creation ones.

\begin{figure*}
\includegraphics[scale=0.75]{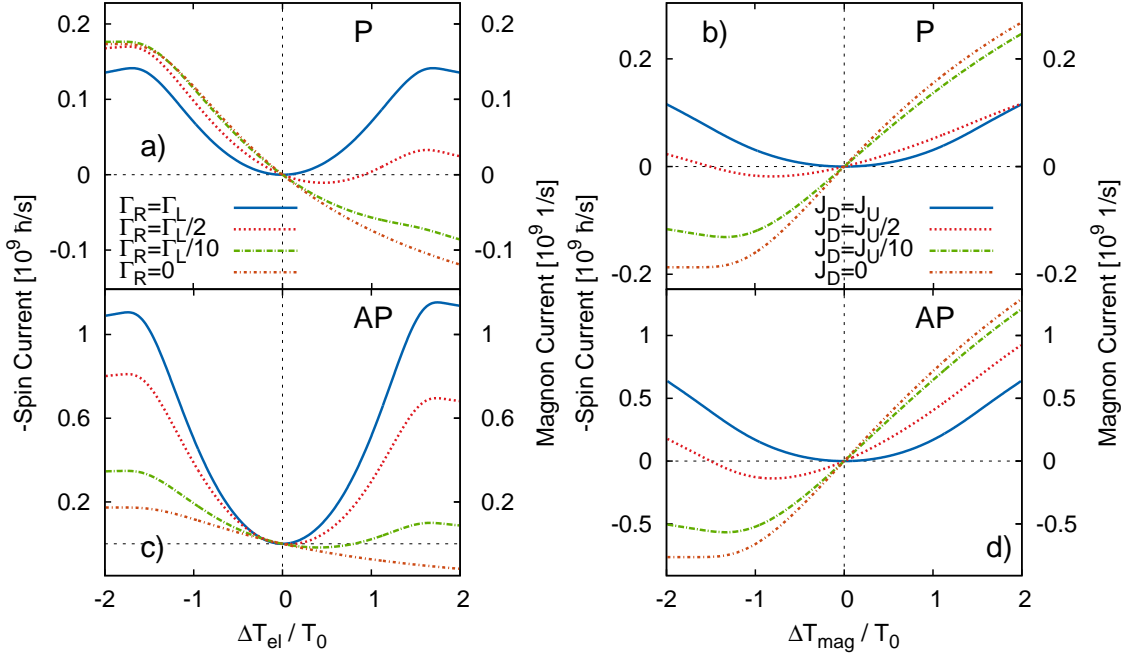}
\caption{\textit{Four-terminal case}. Total spin current $J_{\rm s}^{U}+J_{\rm s}^{D}$ (and the corresponding total magnon current) flowing from the magnonic reservoirs to the electronic ones as a function of the difference in temperatures of the electronic reservoirs, $\Delta T_{\rm el}$, calculated for different values of the coupling parameter $\Gamma_{R}$ (a) and (c) and as a function of the difference in temperatures of the magnonic reservoirs, $\Delta T_{\rm mag}$, calculated for different values of the coupling parameter $J_{D}$ (b) and (d) in the parallel (P) and antiparallel (AP) configuration of the magnetic moments of electronic reservoirs. The other parameters: $\varepsilon_{d}=0$, $g\mu_{B}B=2k_BT_{0}$, $T_{\rm mag}^{0}=T_{\rm el}^{0}=T_{0}$, $p=0.9$, and $k_BT_{0}=0.1$ meV. Apart from this, $\Delta T_{\rm mag}=0$, $\Delta \mu = 0$, $J_{U}=J_{D}=0.1k_BT_{0}$ (a) and (c), and $\Delta T_{\rm el}=0$, $\Delta \mu = 0$, $\Gamma_{L}=\Gamma_{R}=0.1k_BT_{0}$ (b) and (d). }
\label{fig:magnon5}
\end{figure*}

While the magnon current is symmetric with respect to the reversal of thermal bias in the P configuration, a weak asymmetry appears in the AP alignment, compare Figs.~\ref{fig:magnon4}(a) and~\ref{fig:magnon4}(c), where the spin (magnon) current is slightly larger for $\Delta T_{\rm el} >0$. This is especially visible for large polarization factors $p$. The asymmetry is a consequence of the interplay of the difference in the Fermi-Dirac distributions of electrons in the reservoirs and of the asymmetry in the density of states for a nonzero $p$. All this leads to different tunneling rates of spin-$\uparrow$ electrons to the dot (and also different tunneling rates of spin-$\downarrow$ electrons out of the dot) for opposite thermal bias.

Another possibility to convert magnon spin current to an electronic spin current is to apply a finite voltage between the electronic reservoirs.  Figures~\ref{fig:magnon4}(b) and~\ref{fig:magnon4}(d) show the spin and magnon currents flowing from the magnonic reservoirs as a function of the difference in electrochemical potentials of the electronic reservoirs for parallel and antiparallel configurations, respectively. Different curves correspond to indicated values of the polarization factor $p$. For $\Delta \mu =0$, the system is in equilibrium and no current flows through the system. Consider first the  P configuration, where an increase in $|\Delta \mu|$ results in a symmetric and negative magnon current.
We remind that positive (negative) $\Delta \mu$ corresponds to $\mu_R>\mu_L$ ($\mu_R<\mu_L$).
When $|\Delta \mu|$ increases, the probability of electron tunneling to the dot's level $\varepsilon_\downarrow$ effectively increases while tunneling to the $\varepsilon_\uparrow$ level decreases. This imbalance leads to net creation of magnons, and the energy is pumped from the voltage source to the magnonic reservoirs so the magnon current is negative. Again, the magnon current tends to zero in the limit of  $p\to 1$.

In the antiparallel configuration, in turn, a large asymmetry appears with respect to the sign change of the voltage bias.
For positive difference in the electrochemical potentials, $\Delta\mu >0$, the magnon current is relatively large and negative. Moreover, the absolute magnitude of the current increases with increasing polarization factor $p$. To understand this behavior let us consider the limiting situation of $p=1$. If $\Delta\mu >0$
and is relatively large, then only spin-down electrons from the right electrode (in the antiparallel configuration magnetic moment of the right electrode is reversed) can tunnel to the dot, more precisely to the  $\varepsilon_\downarrow$ dot's level. For a sufficiently large $\Delta\mu$, there are no spin-$\uparrow$ electrons that could tunnel to the $\varepsilon_\uparrow$ level, and thus only magnon creation  processes can occur, so the magnon current is negative and relatively large. For smaller values of $p$ or smaller bias voltages, magnon absorption can also occur so the magnitude of magnon current is reduced (though it is negative) as one can clearly see in Fig.~\ref{fig:magnon4}(d) for positive $\Delta\mu$. For negative voltage, in turn, the magnon current is negative for smaller values of $p$, and positive for large values of $p$. Again consider first the case of $p=1$. When $\Delta\mu$ is negative and relatively large, then only spin-$\uparrow$ electrons can tunnel to the dot from the left reservoir. After magnon absorption they change spin orientation and can easily tunnel to the right electrode. The magnon current flows then to the dot and thus is positive. When $p$ decreases, the magnon current also decreases and becomes negative for $p$ larger than a certain value. In the limit of nonmagnetic electronic reservoirs, the current becomes independent of the magnetic configuration and thus is the same as in Fig.~\ref{fig:magnon4}(b), i.e., it is negative.

To achieve a large asymmetry in the thermally-induced magnon current, we introduce different couplings between the dot and the two electronic reservoirs. While the current is symmetric with respect to $\Delta T_{\rm el}$ in a system with symmetrically coupled electronic reservoirs, a significant asymmetry appears for asymmetrical coupling.
Figures~\ref{fig:magnon5}(a) and~\ref{fig:magnon5}(c) show the spin current (and the corresponding magnon current)  flowing from the magnonic reservoirs to the dot  as a function of the temperature  difference $\Delta T_{\rm el}$ for different values of the coupling $\Gamma_{R}$ of the right lead to the dot in the parallel and antiparallel magnetic configurations, respectively. Although, as previously stated, the magnon current in the parallel configuration is smaller than in the antiparallel one, the asymmetry induced by different couplings is larger in the former configuration. Interestingly, for $\Gamma_{R}=\Gamma_{L}/2$ and positive $\Delta T_{\rm el}$, the magnon current disappears also at a point different from $\Delta T_{\rm el}=0$,  and then  becomes negative for smaller values of the parameter $\Gamma_{R}$. There is no sign change of the magnon current for negative $\Delta T_{\rm el}$. Similar asymmetry also appears in  the antiparallel configuration. Note, that in the limit $\Gamma_R=0$ the magnon current is negative for $\Delta T_{\rm el}>0$ and positive for $\Delta T_{\rm el}<0$, and this is true for both P and AP configurations. The system is then effectively three-terminal, but taking into account that both magnonic reservoirs have the same temperature in Figs.~\ref{fig:magnon5}(a) and~\ref{fig:magnon5}(c), one may reduce it further to the M-QD-I system studied in Sec. III. Thus, it is clear that the total spin current due to magnons is transferred to the electronic reservoir (left one in this case).

Figures~\ref{fig:magnon5}(b) and~\ref{fig:magnon5}(d), in turn,  show the spin (magnon) current as a function of the difference in temperatures of the magnonic reservoirs for  indicated values of the coupling  $J_{D}$ between the dot and the  bottom magnonic reservoir and constant value of the coupling of the top magnonic reservoir to the dot.
When the coupling is symmetrical, $J_U=J_D$, the total magnon current flowing to the dot is positive, and symmetric with respect to the sign change  of $\Delta T_{\rm magn}$.
Physically, magnons from the hot magnonic reservoir are pumped to the magnonic reservoir of lower temperature as well as to both electronic reservoirs which also have lower temperature. In the opposite limit of large asymmetry in the coupling, when the bottom magnonic reservoir is weakly coupled to the dot (decoupled in the limit of $J_D\to 0$), the system is effectively equivalent to a three terminal one with two electronic and one magnonic reservoirs. For positive $\Delta T_{\rm mag}$, the magnons flow from the top magnonic reservoir to the electronic ones, while for $\Delta T_{\rm mag}<0$, the magnon flow is reversed, i.e. magnons are created by electrons tunneling from the electronic reservoirs to the dot. The situation is qualitatively similar in both parallel and antiparallel configurations, however, in the antiparallel case the magnon current is remarkably larger.

We note, that when either $J_U=0$ or $J_D=0$, and $\Delta T_{\rm el} =0$,  the four terminal system becomes reduced to the three-terminal one considered in Ref.~\onlinecite{Sothmann}. Attaching one additional magnonic reservoir allows controlling magnonic and electronic spin currents by a temperature difference between the magnonic reservoirs, and by asymmetry in the  coupling of these reservoirs to the dot, as shown in Figs.~\ref{fig:magnon5}(b) and~\ref{fig:magnon5}(d). Further controlling possibility, not considered in Ref.~\onlinecite{Sothmann}, follows from a nonzero $\Delta T_{\rm el}$, see Figs.~\ref{fig:magnon4}(a) and~\ref{fig:magnon4}(c) and Figs.~\ref{fig:magnon5}(a) and~\ref{fig:magnon5}(c), and  from tuning of the dot's energy level and Coulomb coupling parameter, as presented below. 

\subsection{Case of finite $U$}

\begin{figure*}
\includegraphics[scale=0.75]{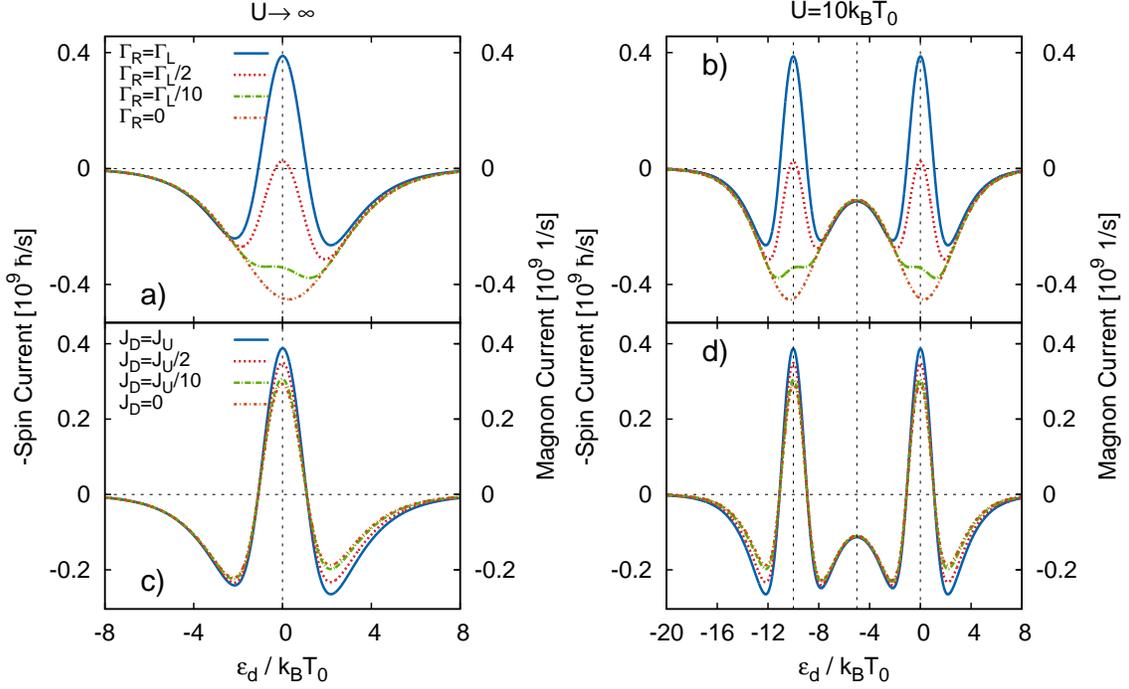}
\caption{\textit{Four-terminal case}. Spin (and magnon) current as a function of dot's energy level $\varepsilon_{d}$, calculated for different values of the coupling parameter $\Gamma_{R}$ (a) and (b), different values of of the coupling parameter $J_{D}$ (b) and (d), and infinite Coulomb interaction, $U\rightarrow\infty$ (left panel) and finite Coulomb interaction, $U=10k_BT_{0}$ (right panel). The other parameters: $J_{U}=J_{D}=\Gamma_{L}=0.1k_BT_{0}$ (a) and (b), $\Gamma_{L}=\Gamma_{R}=J_{U}=0.1k_BT_{0}$ (c) and (d), and $g\mu_{B}B=2k_BT_{0}$, $T_{\rm mag}^{0}=T_{\rm el}^{0}=T_{0}$, $\Delta T_{\rm el}=T_{0}$, $\Delta T_{\rm mag}=0$, $\Delta \mu = 0$, $p=0$, and $k_BT_{0}=0.1$ meV.}
\label{fig:magnon6}
\end{figure*}

Up to now we considered the situation when the bare dot level was located at the Fermi level, while Coulomb interaction was large enough to prevent double occupancy of the dot.
But one of the advantages of quantum dot systems is the possibility of external tuning of its energy levels by a gate voltage, and also external modification of the Coulomb interaction by changing lateral size of the dots. Thus, by external gates one can, in general, modify electronic transport properties of the dots.
 Now we relax the above mentioned assumptions and consider the case when the Hubbard parameter $U$ can be finite (and nonzero), so the double occupancy of the dot is admitted, and the dot's energy level can be tuned externally.
\begin{figure*}
\includegraphics[width=0.75\textwidth]{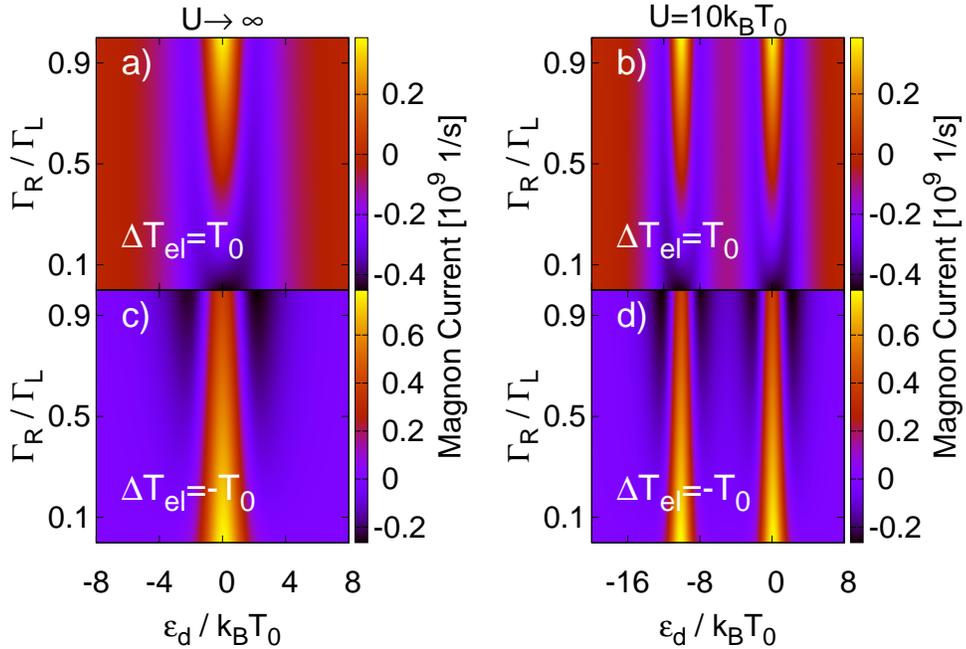}
\caption{\textit{Four-terminal case}. Magnon current as a function of dot's energy level $\varepsilon_{d}$ and coupling parameter $\Gamma_{R}$ for $\Delta T_{\rm el}=T_{0}$ (a) and (b) and  $\Delta T_{\rm el}=-T_{0}$ (c) and (d). Left panel is for infinite Coulomb interaction ($U\rightarrow\infty$), while the right panel is for finite Coulomb interaction, $U= 10k_{B}T_{0}$. The other parameters: $J_{U}=J_{D}=\Gamma_{L}=0.1k_BT_{0}$, $g\mu_BB=2k_BT_{0}$, $T_{\rm mag}^{0}=T_{\rm el}^{0}=T_{0}$, $\Delta T_{\rm mag}=0$, $\Delta \mu = 0$, $p=0$, and $k_BT_{0}=0.1$ meV.}
\label{fig:magnon7}
\end{figure*}

Figures~\ref{fig:magnon6}(a) and~\ref{fig:magnon6}(b) show the spin and magnon currents as a function of the dot's energy level $\varepsilon_{d}$ for infinite as well as finite value of $U$, respectively, and for different values of the coupling parameter $\Gamma_{R}$ between the  dot and right electronic reservoir (while keeping constant coupling between the second electronic reservoir and the dot). Here, both electronic reservoirs are assumed to be non-magnetic, $p=0$.
Consider first the case of $U\rightarrow\infty$ for symmetric coupling, $\Gamma_{R}=\Gamma_{L}$. As follows from Fig.~\ref{fig:magnon6}(a), the magnon current is then positive (magnon absorption processes dominate) for $|\varepsilon_{d}|/k_BT_0 <1$. In turn, when  $|\varepsilon_{d}|/k_BT_0 >1$, the magnon emission processes become dominant and the magnon current is negative. This is because both spin-up and spin-down dot's levels are above (or below) the Fermi level so electrons from the right electronic reservoir  cannot take part in transport due to its low temperature. The current is then governed by the left reservoir which has much higher temperature, and since its temperature is also higher than that of the magnonic reservoirs, energy (magnons) can flow to the magnonic reservoirs and the magnon current is negative.
\begin{figure*}
\includegraphics[scale=0.75]{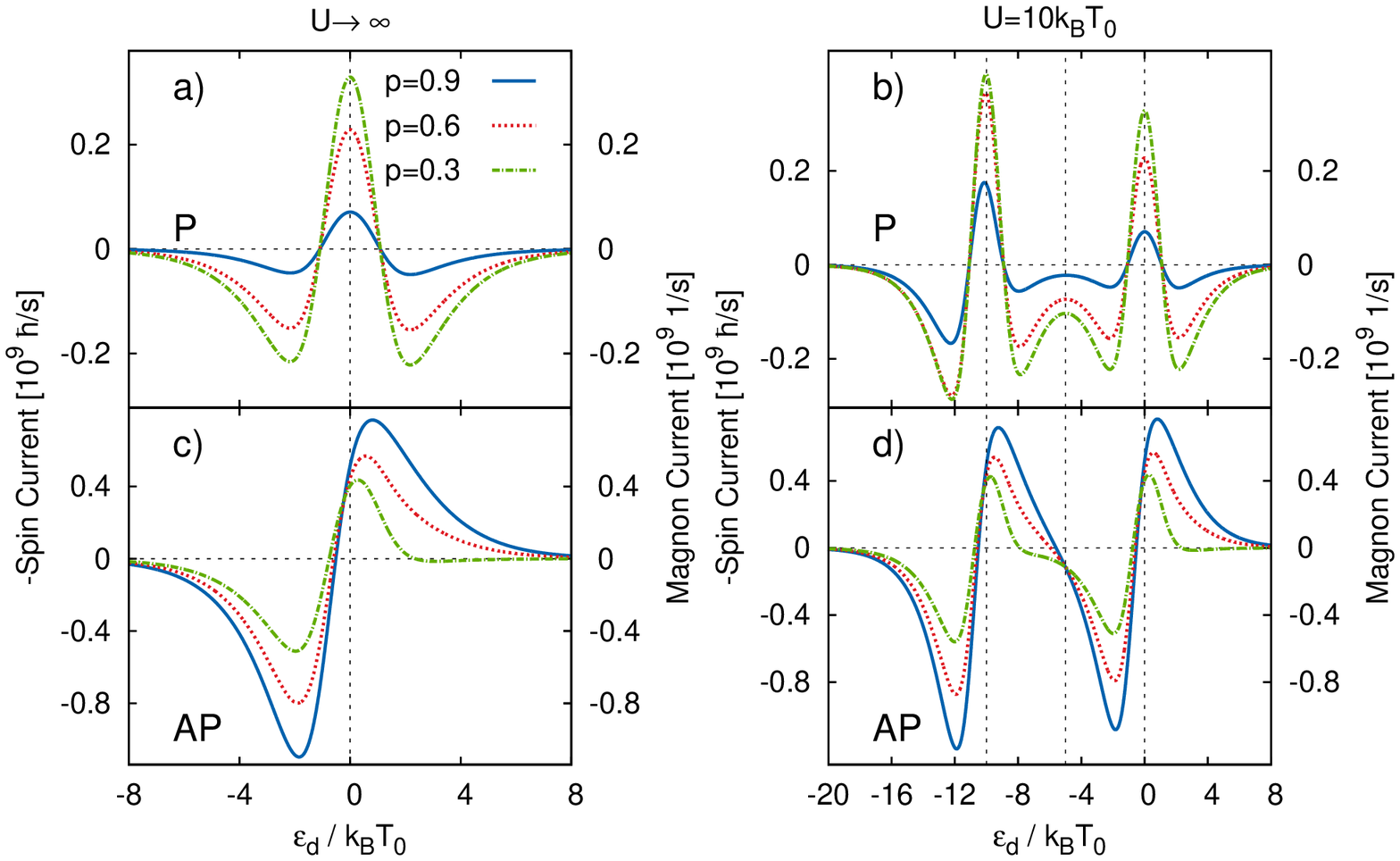}
\caption{\textit{Four-terminal case}. Magnon current as a function of the dot's energy level $\varepsilon_{d}$ for indicated values of the polarization factor $p$, and  for parallel (P) (a) and (b) and antiparallel (AP) (c) and (d) magnetic configurations. Left panel is for infinite Coulomb interaction ($U\rightarrow\infty$), while the right panel is for finite Coulomb interaction, $U=10k_{B}T_{0}$. The other parameters: $J_{U}=J_{D}=\Gamma_{L}=\Gamma_{R}=0.1k_BT_{0}$, $g\mu_BB=2k_BT_{0}$, $T_{\rm mag}^{0}=T_{\rm el}^{0}=T_{0}$, $\Delta T_{\rm mag}=0$, $\Delta T_{\rm el}=T_{0}$, $\Delta \mu = 0$, and $k_BT_{0}=0.1$ meV.}
\label{fig:magnon8}
\end{figure*}
When decreasing the coupling parameter $\Gamma_{R}$ by a factor of 2, one finds almost complete suppression of the magnon current at $\varepsilon_{d}=0$. Further decrease in the coupling parameter $\Gamma_{R}$ results in a sign change of the magnon current, whose absolute magnitude then increases with a further decrease in $\Gamma_{R}$. When coupling to the right electronic reservoir is negligible (note this reservoir has the lowest temperature), then the magnon current is negative independently of $\varepsilon_{d}$. The system is then equivalent to a magnon reservoir connected to an electronic one and it is clear that energy (magnons) should flow from the reservoir of higher temperature  to that of the lower temperature, i.e. from the electronic to magnonic reservoirs.
Finite $U$ [Fig.~\ref{fig:magnon6}(b)] leads to the appearance of another peak for $\varepsilon_{d}=-U$, which can be considered as a Coulomb counterpart of the main peak. Due to the particle-hole symmetry, the whole spectrum is also symmetric.

Figures~\ref{fig:magnon6}(c) and~\ref{fig:magnon6}(d) show the results similar to those discussed above, but for asymmetric coupling of the magnonic reservoirs to the dot. Increasing this asymmetry results in a slight decrease in the magnon current near $\varepsilon_{d}=0$,  and also near $\varepsilon_{d}=-U$ in the case of finite $U$. A similar decrease in the magnitude of the magnon current also appears near $\varepsilon_d/k_BT_0\approx 2$ and in the corresponding region in the case of finite $U$.

Behavior of the  magnon current (and also of the spin current) with the dot's level energy and reduced coupling to one of the reservoirs (either electronic or magnonic), was shown in Fig.~\ref{fig:magnon6} for the situation when the temperature of the electronic reservoir, whose coupling to the dot was constant,  was the largest temperature in the system. It is interesting to look how this behavior changes when this temperature will be the lowest one.  This is presented in Fig.~\ref{fig:magnon7}, where the magnon current is shown as a function of the dot's energy level and coupling of the right electronic reservoir for $\Delta T_{\rm el}=T_0$ [Figs.~\ref{fig:magnon7}(a) and~\ref{fig:magnon7}(b)] and for $\Delta T_{\rm el}=-T_0$.
For $\varepsilon_{d}=0$ and $\Gamma_{R}\ll\Gamma_{L}$, the magnon current is negative for $\Delta T_{\rm el}=T_0$, and positive for $\Delta T_{\rm el}=-T_0$. This behavior is obvious from the thermodynamic point of view (energy flows from the reservoir of higher temperature to that of lower temperature). Increasing $\Gamma_{R}$ results in splitting of the peak into two negative peaks and appearance of a single positive peak for $\Delta T_{\rm el}=T_0$,
and reduction of the positive peak and appearance of two negative peaks for $\Delta T_{\rm el}=-T_0$. Note, for $\Gamma_R/\Gamma_L\to 1$, the magnon current is the same for $\Delta T_{\rm el}=T_0$ and $\Delta T_{\rm el}=-T_0$. Similar features are exhibited also by the second peak in the case of finite $U$.

Figure~\ref{fig:magnon8} presents magnon current as a function of dot's energy level $\varepsilon_{d}$ for different values of the spin polarization factor $p$, and  for both parallel and antiparallel magnetic configurations of the electronic reservoirs. Figure~\ref{fig:magnon8}(a) clearly indicates that the magnon current in the parallel configuration decreases with increasing $p$. Origin of this behavior was already discussed above.
In the antiparallel configuration, on the other hand, the effect of polarization is opposite, i.e. the magnon current increases with increasing polarization factor $p$. Apart from this, the current is asymmetric with respect to position of the dot's level.
Figures~\ref{fig:magnon8}(b) and~\ref{fig:magnon8}(d) show similar behavior in the case of finite Hubbard parameter $U$.

\section{Four-terminal case: charge current}

\subsection{Conversion of magnon current to charge current}

Above we analyzed the conversion of magnon current flowing from the magnonic reservoirs to spin current in the electronic leads. However, the magnon current can also generate  a charge current, as shown in Ref.~\onlinecite{Sothmann} for a three-terminal case. To show this in the system under consideration we assume no voltage and no temperature difference
between the  electronic reservoirs, $T_L=T_R$ and $\mu_L=\mu_R$, and continuously
vary the temperature difference between the magnonic reservoirs.

In the parallel magnetic configuration of the electronic reservoirs, the charge current flowing between these reservoirs
vanishes, because the rates of electron tunneling
to and out of a given electronic reservoir are the same.
More precisely, the flux of electrons with a given spin orientation is compensated by the
flux of electrons with opposite spin and flowing in the opposite direction. As a result, a pure spin
current can flow,  with no accompanying charge current. This holds also in the presence of
some asymmetry in the couplings of the dot to the left and right electronic reservoirs.
However, the situation changes in the antiparallel magnetic configuration, where a
finite charge current can flow.

Let us briefly explain the mechanism of charge current generation  by a magnon current in the antiparallel  configuration,
assuming first half metallic electronic reservoirs,  $p=1$.
When the dot is initially empty, then an electron with spin-up orientation
can tunnel from the left electronic reservoir to the dot.
Upon spin reversal by absorbing a magnon, this electron can tunnel further to
the right ferromagnetic lead. As a result, a flux of electrons flows from the left to right electronic lead.
This charge current assists the magnon current flowing from the hotter magnonic reservoir.
There is also a flux of electrons flowing from the right to left electronic reservoirs. Now,
a spin-down electron tunnels from the  right reservoir to the dot, then it creates  a magnon in the magnonic reservoirs and
as a spin-up electron tunnels to the left reservoir.
However, the tunneling rate of such processes is much reduced in comparison to the tunneling rate
of electrons flowing from the left to right lead, as already discussed above in the case of magnon current.

\begin{figure}
\includegraphics[width=0.9\columnwidth]{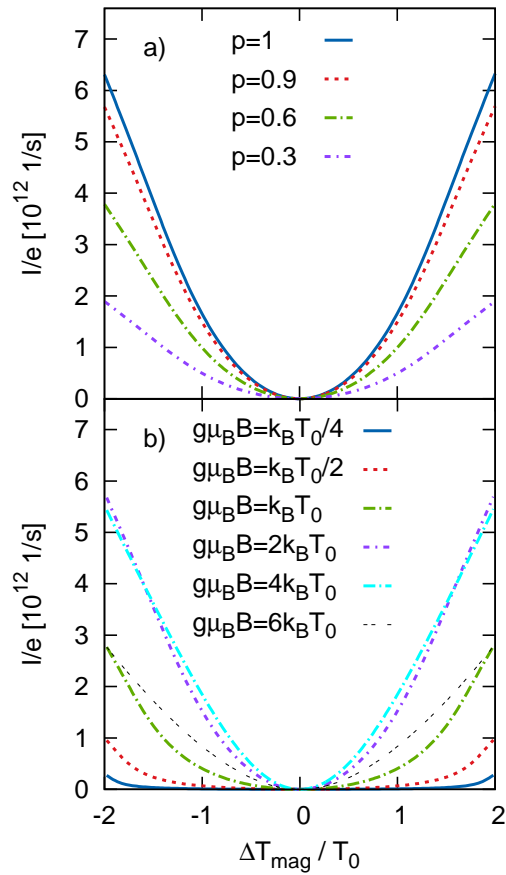}
%[scale=0.6]
\caption{\textit{Four-terminal case}. Electron (particle) current in the antiparallel magnetic configuration of the electronic reservoirs. The current is plotted  as a function of the difference in temperatures of the magnonic reservoirs, $\Delta T_{\rm mag}$, and is calculated for indicated values of the polarization factor $p$ and  $g\mu_BB=2k_BT_0$ (a), and for indicated values of magnetic field $B$ and $p=0.9$ (b). The other parameters: $\varepsilon_{d}=0$, $T_{\rm mag}^{0}=T_{\rm el}^{0}=T_{0}$, $\Delta T_{\rm el}=0$, $\Delta \mu = 0$, $J_{U}=J_{D}=0.1k_BT_{0}$,  $\Gamma_{L}=\Gamma_{R}=0.1k_BT_{0}$, and $k_BT_{0}=0.1$ meV.   }
\label{Fig:delMp}
\end{figure}

In Fig.~\ref{Fig:delMp}(a) we show the charge current normalized to electron charge, which effectively is the particle (electron) current, flowing from the left to right electronic reservoir
in the antiparallel  magnetic configuration (note that the charge
current has opposite sign to the particle current). The current is plotted there as a function of the difference $\Delta T_{\textrm{mag}}$
in temperatures of the magnonic reservoirs  for indicated values of the spin polarization factor $p$ and for
$\Delta T_{\textrm{el}}=0$, $\Delta\mu=0$, and equal couplings of the magnonic reservoirs to the dot.
The electron current is then symmetric with respect to the reversal of the thermal bias and  positive as electrons flow from the left to right lead
(see explanation above).
No current flows for $\Delta T_{\textrm{mag}}=0$, and
the current grows monotonically with increasing  $|\Delta T_{\textrm{mag}}|$.
From Fig.~\ref{Fig:delMp}(a) follows that the current drops with decreasing spin-polarization factor $p$.
The dominant contribution to the current comes from spin-up electrons  tunneling to the dot from the left metallic lead.
For $p<1$, also spin-down electrons from the left metallic lead and spin-up electrons from the right lead contribute to current.
Taking all these tunneling processes into account, one can conclude
that the particle flow from left to right is partly compensated by that from right to left, so the particle current becomes reduced with decreasing spin polarization factor, and vanishes  for  $p=0$.
The current also depends on the magnetic field which filters the magnons passing through the dot.

In Fig.~\ref{Fig:delMp}(b) the current is shown as a function of $\Delta T_{\textrm{mag}}$
for several values of magnetic field.
Although, the magnon current achieves significant magnitude for small magnetic fields $B$ (see preceding sections),
the resulting electron current becomes then strongly suppressed. A small magnetic field leads to small Zeeman
splitting of the dot's energy level, and thus both $\varepsilon_{\uparrow}$ and $\varepsilon_{\downarrow}$
are located near the Fermi level of the electronic reservoirs: $\varepsilon_{\uparrow}$ is slightly below the Fermi level, whereas $\varepsilon_{\downarrow}$ is slightly above the Fermi level. Thus, the tunneling rates
of electrons flowing from the right to left metallic leads becomes enhanced. As described above, these tunneling events
compete with the (dominant) tunneling processes from the left to right reservoirs. As a result, the net current becomes suppressed.
When magnetic field increases, the current initially grows up to a certain value of $B$, and then it becomes reduced with a further increase in
magnetic field. For sufficiently large magnetic fields, tunneling processes from the right to left electronic reservoir become
strongly suppressed and practically do not contribute to the current. Thus, one might expect an increase in current.
However, the magnon current decreases with increasing magnetic field due to a small population of high energy magnons.
Competition of these two effects leads to a nonmonotonous
dependence of the current on applied magnetic field, and to suppression of current for large values of the magnetic field $B$.

\begin{figure}
\includegraphics[width=0.9\columnwidth]{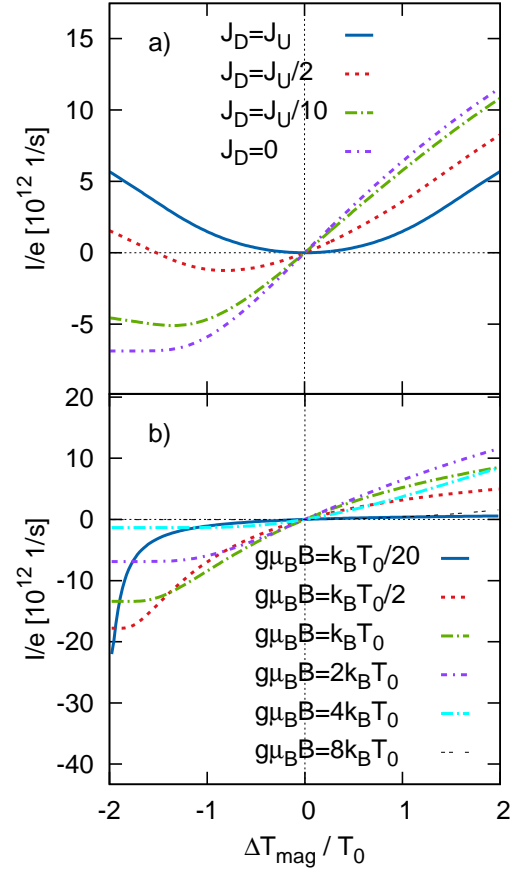}
%[scale=0.6]
\caption{\textit{Four-terminal case}. Electron (particle) current in the antiparallel magnetic configuration of the electronic reservoirs. The current is plotted as a function of the difference in temperatures of the magnonic reservoirs, $\Delta T_{\rm mag}$, and is calculated for indicated values of the coupling parameter $J_D$ and  $g\mu_BB=2k_BT_0$ (a), and for indicated values of magnetic field $B$ and  $J_D=0$ (b). The other parameters: $\varepsilon_{d}=0$, $T_{\rm mag}^{0}=T_{\rm el}^{0}=T_{0}$, $p=0.9$, $\Delta T_{\rm el}=0$, $\Delta \mu = 0$, $J_{U}=J_{D}=0.1k_BT_{0}$, $\Gamma_{L}=\Gamma_{R}=0.1k_BT_{0}$, and $k_BT_{0}=0.1$ meV. }
\label{Fig:delTmJ}
\end{figure}

In Fig.~\ref{Fig:delTmJ}(a) we show electron current when the coupling to one of the magnonic reservoir becomes reduced, whereas coupling to the other one
is kept constant. When both couplings to the magnonic reservoirs are the same ($J_U=J_D$), the current is positive and symmetric with respect to the
temperature difference $\Delta T_{\textrm{mag}}$, and flows from the left to right reservoirs  in the whole range of $\Delta T_{\textrm{mag}}$. When the coupling to one of magnonic reservoirs (let's say $J_D$) is decreased, the current becomes asymmetric and for a significant asymmetry in the couplings it becomes negative for $\Delta T_{mag}<0$. Let us consider in more details the limiting case of $J_D=0$, i.e. when the magnonic reservoir corresponding to $\alpha =D$  becomes decoupled from the dot. When $\Delta T_{\textrm{mag}}>0$, the magnons flow to the dot and the electron current is positive, i.e. electrons flow from the left to right electronic reservoir. In turn, for $\Delta T_{\textrm{mag}}<0$, the electron current flows in the opposite direction and magnons are excited in the  magnonic reservoir.
It is also worth noting that for a not too large asymmetry in the couplings to the magnonic reservoirs, the electron current vanishes for a certain nonzero
value of $\Delta T_{\textrm{mag}}$.

Interestingly, by tuning the magnetic field one can obtain a strong thermal rectification of electron current in the three-terminal setup, $J_D=0$. To show this we plot in Fig.~\ref{Fig:delTmJ}b  the electron current for indicated values of the magnetic field. The strongest rectification occurs for relatively small magnetic fields. For $g\mu_BB=k_BT_0$ the electron current becomes strongly suppressed when the temperature of the magnonic reservoir is higher that the temperature of metallic leads, $\Delta T_{\textrm{mag}}>0$, whereas for $\Delta T_{\textrm{mag}}<0$ it achieves a relatively large value for large difference in temperatures of the magnonic and electronic  reservoirs.
Generally, the electron current depends in a nonmonotonic way on the magnetic field. For $\Delta T_{\textrm{mag}}>0$, the electron current is positive and
grows initially with increasing  magnetic field (up to certain value of magnetic field), and then decreases with a further increase in $B$.
This initial increase of electron current  can be understood when taking into account the fact that electron current depends on the Zeeman splitting of the dot's level. As explained above, the flux of electrons flowing from right to left becomes significant for small values of the magnetic field. Upon reaching a  maximum value, the current  decreases with a further increase in $B$, and this behavior follows directly from the Bose-Einstein distribution of magnons,  as already mentioned above (population of magnons with large energy is small, which results in a small magnon current, and thus low electron current).
Similar behavior also holds for negative $\Delta T_{\textrm{mag}}$, except a small region of  large negative values of $\Delta T_{\textrm{mag}}$, where the absolute value of electron current  decreases monotonically with increasing magnetic field. Apart from this, the electron  current is negative as the tunneling processes are now assisted by creation of magnons in the magnonic reservoirs.

\begin{figure}
\begin{center}
\includegraphics[width=0.9\columnwidth]{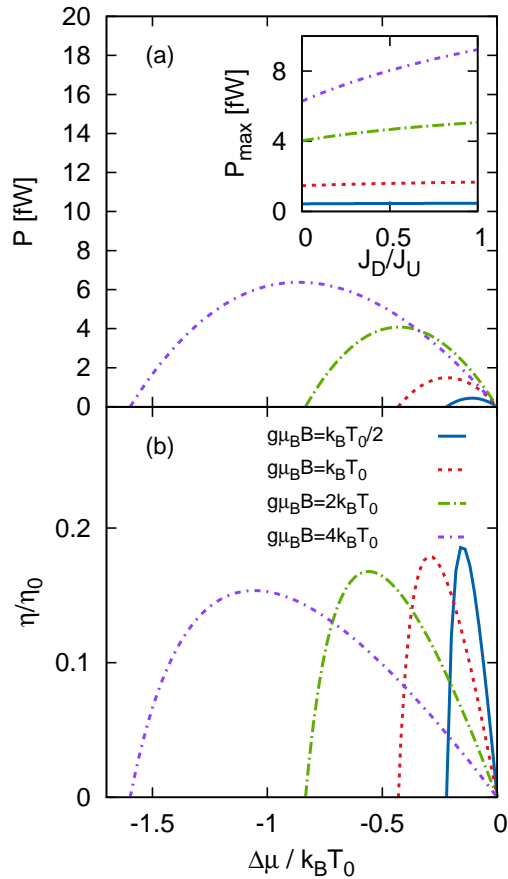}
  \caption{\label{Fig:powerB}
  (a) Power as a function of difference in the electrochemical potentials $\Delta \mu=eV$ calculated for indicated values of magnetic field $B$ and for $p=0.9$. (b) The corresponding efficiency $\eta$ normalized to Carnot efficiency $\eta_{0}$.  The other parameters: $U\rightarrow \infty$, $J_{D}=0$, $p=0.9$, $T_{\textrm{mag}}=2T_{0}$, $T_{\textrm{el}}=T_{0}$, $J_U=0.1k_BT_0$, and $k_BT_{0}=0.1$ meV. The inset shows the maximal power $P_{\textrm{max}}$ as a function of the coupling of the dot to the bottom lead $J_{D}$ for indicated values of magnetic field $B$.
  }
  \end{center}
\end{figure}

\subsection{Heat engine}

As shown by Sothmann and B\"uttitker~\cite{Sothmann} for a three-terminal device, the system can work as a nanoscale thermoelectric heat engine with an output power. Below we consider magnetic field dependence of the corresponding characteristics. 
To do this we assume a finite bias voltage $eV=\Delta \mu$ against
which the thermoelectric current can do some work. Furthermore, we assume no temperature difference between the electronic reservoirs and no temperature difference between the magnonic reservoirs, $\Delta T_{\textrm{mag}}=0$ and $\Delta T_{\textrm{el}}=0$. However, the magnonic and electronic reservoirs have different temperatures. The output power is given by $P=eIV$. Assuming $T^0_{mag}>T^0_{el}$ ($T_U=T_D=T^0_{mag}$ and $T_L=T_R=T^0_{el}$) the efficiency of the device is given by $\eta=P/J^Q$ with $J^Q=J^Q_U+J^Q_D$. Here, $J^Q_\alpha$ ($\alpha=U,D$) denotes heat current flowing out of the $\alpha$-th magnonic reservoir. In turn, for $T^0_{mag}<T^0_{el}$ the heat current becomes $J^Q=J^Q_L+J^Q_R$ with $J^Q_\beta$ ($\beta=L,R$) denoting heat current flowing out of $\beta$-th metallic leads.

As no charge current flows in the parallel magnetic configuration, we study only the antiparallel one, and consider first the three terminal device in which one of the magnonic reservoir is decoupled from the dot, e.g.  $J_D=0$. In Fig.~\ref{Fig:powerB} we show the output power and efficiency of the device as a function of the external bias voltage for indicated values of applied magnetic field $B$. The output power grows with increasing $V$ until it reaches a maximum. The power decreases with a further increase in $V$ and finally achieves zero at the voltage $V=V_s$, at which the thermoelectric current becomes totaly compensated by the current induced due to applied bias voltage. The bias voltage at which the current is suppressed to zero corresponds to the Seebeck voltage induced by a difference in temperature under open circuit operation. The power can be extracted only for $V\in(0,V_s)$.

Generally, the maximum of the power increases with increasing magnetic field $B$ (at least for indicated values of $B$) and the absolute value of the thermoelectric  voltage also grows. One can also notice that for small values of $V$ the power follows a nonmonotonic filed dependence of the electron current studied in the previous section. However, for larger $V$ the power seems to grow monotonically with magnetic field $B$. It turns out that for larger magnetic fields (not shown) the maximum power decreases following the nonmonotonic dependence of the current as described earlier. Including the fourth terminal (second magnonic reservoir), results in a greater maximum power as shown in the inset in Fig. ~\ref{Fig:powerB}(a).
In Fig. ~\ref{Fig:powerB}(b) the efficiency of heat to work conversion is shown as a function of applied bias voltage. The efficiency vanishes, similarly  as the power, at zero voltage and at $V_s$. In the former case, $\eta=0$ as there is no external bias voltage applied ($V=0$), whereas in the latter case power vanishes, and thus $\eta=0$ as a result of no current flowing. It is interesting to note that although the maximum power reveals rather strong dependence on the magnetic field, the maximum efficiency becomes only weakly dependent on $B$.
Moreover, the maximum efficiency does not occur at the same  voltage as the maximum power, in general.

\section{Summary and conclusions}

We have considered a single-level quantum dot connected in a general case to two electronic (spin-polarized) and two magnonic reservoirs.
Some specific situations, when only two magnonic, or one magnonic and one electronic reservoirs are coupled to the dot have been also analyzed.
The main objective of this analysis was focused on the conversion of spin current carried by magnons to spin current of electronic origin, and {\it vice versa}.
In a general situation, all the  reservoirs have different temperatures and the spin and charge currents are driven either by an external voltage applied to the electronic reservoirs and by difference in temperatures of the reservoirs. In the case of infinite $U$, equal temperatures of the electronic reservoirs, and vanishing coupling of one of the magnonic reservoirs to the dot, the system reduces to that considered by Sothmann and B\"uttitker.\cite{Sothmann}

Employing the master equation to describe stationary tunneling  rates from the reservoirs to the quantum dot, we have calculated both spin current and magnon current.
We have shown that using specific arrangements of various reservoirs, one can control the spin current flowing in the system, and thus also the conversion of magnon current to electronic spin current, either by an external voltage or by temperatures of the reservoirs. Moreover, the magnon current conversion can also be tuned by an external gate voltage which shifts position of the dot's level. The quantum dot plays thus a role of a convertor of magnon current to spin current and {\it vice versa}. We have also considered the influence of the Coulomb interaction in the dot on the magnon current and its conversion to electronic spin current.
Finally, we have studied the conversion of magnon current to charge current, and calculated the power and efficiency of a heat engine based on the  system under consideration.

\begin{acknowledgments}
This work was supported by National Science Centre in Poland as Project No. DEC-2012/04/A/ST3/00372.
\end{acknowledgments}

\end{document}